\documentclass[aps,prd,superscriptaddress,eqsecnum,amsfonts,showpacs,epsfig]{revtex4}
\usepackage{epsfig}

\newcommand{\be}{\begin{equation}}
\newcommand{\ee}{\end{equation}}
\newcommand{\bea}{\begin{eqnarray}}
\newcommand{\eea}{\end{eqnarray}}

\newcommand{\nn}{\nonumber}
\newcommand{\ep}{i\epsilon}
\newcommand{\om}{\omega}

\begin{document}


\title{Timelike behavior of the pion electromagnetic form factor in the functional formalism. }

\author{V. \v{S}auli}

\email{sauli@ujf.cas.cz}
\affiliation{Department of Theoretical Physics, Institute of Nuclear Physics, \v{R}e\v{z} near Prague, CAS, Czech Republic  }

\begin{abstract}
The electromagnetic form factor of the pion is calculated within the use of functional formalism. We develop integral representation for  the
minimal set of  Standard Model Green's functions and  derive the dispersion relation for the form factor  in the two flavor 
QCD isospin limit $m_u=m_d$.
We use the dressed quark propagator as obtained form the gap equation  in Minkowski space and within the Dyson-Schwinger  equations formalism to derive the approximate dispersion 
relation for the form factor for the first time. We evaluate the form factor for the spacelike as well as for the timelike momentum in the presented formalism.
 A new Nakanishi-like form of integral representation is proved on the basis of the vector Bethe-Salpeter equation for the quark-photon vector with
a ladder-rainbow kernel. The Gauge Technique turns out to be a  part of the entire structure of the vertex.
 In the analytic approach presented here, it is shown  that  the a large amount of the  $\rho$-meson peak in the cross section
 $e^+e^-\rightarrow \pi^+\pi^-$ is governed by the gauge invariance of QED/QCD, i.e.  by  Gauge Technique constructed quark-photon vertex.
This approximation naturally explains  the  broad shape of the  $\rho$-meson peak.
\end{abstract}

\maketitle

\section{Introduction}

The understanding of QCD as a quantum field theory would not be complete without mastering  all  domains-- the perturbative calculation of  high-energy  processes, as well as by achieving  success in  nonperturbative evaluations of low-energy  hadronic production processes. In the latter case the timelike character of  transferred momenta is an additional challenge.  Developing a new nonperturbative  technique opens new  prospects in this last and not yet well theoretically mastered area.  In this respect, the   charged meson form factors  and the transition meson form factors constitute rather precise  data on the electromagnetic structure of the light meson. Simultaneously, both processes  are  simple enough for theoretical description based on quark and gluon degrees of freedom.   
 The pion form factor $F(s)$ in a timelike region of momenta ($s=q^2>0$) carries nontrivial information in  amplitudes for the production of the two lightest hadrons: a $\pi_+\pi_-$ pair. It has been  measured with unprecedented accuracy 
($0.5\%$ \cite{I})  in the domain of   the appearance of a striking resonant structure. Using a phenomenological description based on Breit-Wigner fits, the resonant structure corresponds
to a  775 MeV heavy   $\rho$ meson, where  very nearby at $s=780$ MeV  a small ($\simeq 5\%$) admixture of very narrow $\omega$ resonance is also observed \cite{COPETH1997}. The physical neutral $\rho$,
 having  charged partners, is hence effectively considered as a neutral component of the vector isotriplet, while the physical $\omega$ is, within a good approximation, a strong isosinglet.
Moreover, vector  mesons could be practically untouched by the QCD non-Abelian anomaly, and both aforementioned light electrically  
neutral vectors could be practically identical in the isospin limit defined as $m_u=m_d$. However, light vector meson  couplings to other mesons are quite different, providing their cross-section shapes differ very dramatically in various processes. 
We brought some new hints, supported by  quantitative results, which offer   the fact that various components of QCD/QED vertices contribute very differently in different processes. More concretely, facing the amount of the pion peak obtained, it is very likely that seven out of the eight components   of transverse quark-photon vertices could contribute by only a limited amount, suppressed  in vicinity of $\rho$-meson peak and its shape is dictated by the gauge invariance more then we expected.

In a spacelike region $[F(t),t=-q^2; t>0]$,  the available data \cite{data2006,tade2007,aubert2007,data2007,BABAR2012}  are represented by a smooth decreasing curve for which
the perturbative QCD prediction \cite{CHZH1977,FAJA1979,LEBR1979,EFRA1980,LE2002} reads
\be \label{QCDPT}
F_{\pi}(t)\rightarrow \frac{64\pi^2f_{\pi}^2}{(11-2/3n_f)\, t\,  L(t) }\left[1+B_2 L_t^{-\frac{50}{581}}+B_4 L_t^{-\frac{364}{405}}+O(L_t^{-1})\right]^2 
\ee
where  $L_t=\ln{(t/\Lambda^2_{QCD})}$ and the coefficients $B_i$ are related to the nonperturbative part-- the pion distribution and the light-cone Bethe-Salpeter wave function.
The actual  asymptotic predictions  within today's available experimental range  $Q^2 \simeq $100 GeV$^2$ moves the validity of perturbative QCD predictions
 more toward the deep spacelike scale.

New interesting resonant structures, e.g., the deep dip at 1.5 GeV, and other heavier resonances have been found in the shape of $F$ by using the Initial Photon State method  in the BABAR 2012 experiment \cite{BABAR2012}.
With increasing energy, a  theory like vector meson dominance  rapidly becomes a  tautology of what is observed in the experiment: the experimental  masses of ground state and excited mesons become  mass parameters  of the theory,
while the widths of resonances very much reflect the introduced effective couplings among various meson. Chiral perturbation theory \cite{GALE1985}   has  calculated the electromagnetic form factors  near the threshold in various approximations 
\cite{CFU1996,BCT1998,BITA2002,BIJDHO2003,KAROL2012,DGL1990,GAME1991,DN1997,CA471,PIPO2001,OOP2001,TY2002}, while the evaluation  at higher energy, $Q>0.5 GeV$ is  out of  convergence
with the theory and further phenomenological  degrees of freedom need to be added in order to continue to higher energy \cite{DGKST2015}. 
The functional approach  provides good results for spacelike mesonic form factors\cite{MT99,MATA200a,MATA200b,kacka2005,CHA2013,RCBRG2016,CDCL2017,EFWW2017,WEFW2017,DRBBCCR2019,YPNFS2021}, 
where the approach  connects all lengths naturally: it is nonperturbative at low $Q^2$ where QCD is strong and it complies with  perturbation theory at spacelike asymptotic.  Due to known limitations and obstacles, only a few studies \cite{VHP2020,TPF2020,LSA2021}  based on the quantum field theory functional formalism offer a result for the function $F$  in the entire Minkowski space. A new approach \cite{YPNFS2021}  for the evaluation of $F$  based on the integral representation of Bethe-Salpeter functions is employed at level of the constituent quark model (in this approximation , the running of quark masses as well as the momentum dependence of the quark  renormalization function is ignored) and the authors restrict themselves to only the spacelike argument of photon momenta. In the presented study, we follow similar lines as the authors in Ref. \cite{YPNFS2021}, 
but we take the momentum dependence in the quark propagator into account. Consequently, for the first time, we calculate the electromagnetic pion form factor in the entire Minkowski space.

The  distinct  shapes  of $\rho$ and $\omega$ resonances  appearing in processes where they dominate  [say, the former in the function $F(s)$ 
and the latter in the $3\pi$ production]
  are a known striking feature. The $\rho$-meson is a broad resonance, while  the $\omega$ peak is 20 times more narrow.
  The 70 $\% $ contribution of the two pion production  cross section $\sigma(ee\rightarrow \pi\pi)$   to the muon anomalous magnetic
 momentum $a_{\mu}$ is an integral quantitative  expression of the above statement. Single pion or three pions productions dominated by the exchange of $\omega$ (and not $\rho$) mesons  in $e^+e^-$ collisions contribute to $a_{\mu}$ by a remarkably smaller amount.  
To explain this,   new terms with new  couplings related with $\rho-\omega-\pi $ mixing  are incorporated in the effective theories of QCD \cite{KKW1996,LL2009,B2009,TSLE2013,DGKST2015} . 
These  new effective couplings ensure the tree level decay of both mesons: the decay of  $\rho\rightarrow \pi\pi$ happens at a point, while the  decay of $\omega$ 
happens  through the radiation of pions and the subsequent decay of 
virtual  $\rho\rightarrow \pi\pi$, so $\rho$  participates virtually and its  propagation   slows the decay of the $\omega$ meson.
Although less effective and more demanding in practice, it is also worthwhile to understand this origin  from a microscopic explanation based on the quark and gluon degrees of freedom.   An understanding of the detailed shape of $\rho$-meson resonance with no more than QCD Lagrangian parameters is certainly not equivalent to an
 empirical introduction of different phenomenological couplings between light vectors and pseudoscalars.

   It is useful  \cite{DGL1990,GAME1991,DN1997,CA471,PIPO2001,OOP2001,TY2002,LE2002,BEBUDULI2011,HKLNS2014,GOGUAD2012,SORO2019} to consider  
   the pion form factor  as the boundary value of an analytical function
   which has  a cut on the timelike axis of the $q^2$ variable, which  starts in the branch point $s_{th}=q_{th}^2=4m_{\pi}^2$, the production threshold.
  Thus the electromagnetic form factor $F_{\pi}(t)$ and the production form factor $F(s)$ can be evaluated from the   dispersion relation for $F$:
  \be  \label{DR}
  F(q^2)=\int_0^{\infty} d\omega \frac{g(\omega)}{q^2-\omega+\ep} \, ,
  \ee
with the unique spectral function $g$, which represents the imaginary  part of $F(s)$ itself, $\Im F(s)=-\pi g(s)$ and which vanishes below $s_{th}$, provided $F(s)$  is the real 
function there. We will simply  write $F_{\pi}(x)$ for any momentum either for spacelike $x<0$ or for the timelike argument   $x=s>0$ and  for the  Euclidean scalar product we have   $q_E^2=-t$  in the convention used in this paper.

We present the technique, which within the use of quark and gluon degrees of freedom,  leads  to the form of the dispersion relation in Eq. (\ref{DR}).
It does not use predetermined properties of   vector mesons; they appear as a solution of Schwinger-Dyson equations for propagators and vertices.
Vector meson masses are not an input anymore; furthermore the $\rho$-meson is not taken as a  stable hadron- it has  no associated real pole in S-matrix and 
therefore it does not come from the solution of the homogeneous bound state BSE at all. 
The function $g$ in  Eq. (\ref{DR}) is then given by  a multidimensional integral over the spectral functions of quark propagators and  Nakanishi weight functions for the Bethe-Salpeter pion vertex function as well 
as over the weight functions which appear in the Integral Representation (IR) for the  quark-photon vertex.

 We will report on an exploratory study of the
time-like pion electromagnetic form factor using  IRs of QCD Green's functions, which 
are derived from nonperturbative  truncation of QCD/QED  Dyson-Schwinger Equations (DSEs).
The IR is introduced in the Sec. IV, and the proof is relegated  to the  appendices.
 Proposed IR for vertices instantly  offer  the analytical continuation of Euclidean solutions for  QCD Green's functions,
 as well as for hadronic form factors. 
  In order to get  the necessary functions for calculation of the pion form factor $F$, we use the solution of a combination of DSEs and the  Bethe-Salpeter equation (BSE),  which was employed recently for the purpose of calculation 
of the pion transition form factor \cite{TPF2020} and hadron vacuum polarization \cite{VHP2020}.
Furthermore,  we derive a formula  for the form factor $F$ in this limit and calculate  the integral in Eq.  (\ref{DR}) numerically.

 
\section{The electromagnetic pion form factor for timelike argument and the minimal set of equations of motions}

The evaluation of the pion form factor is a typical quantum field theory problem which involves bound states as  final or initial states.
How to calculate such a transition in the BSE approach  is generally known \cite{MANDEL}. Since we deal with gauge theory, which  has the additional approximate  global symmetries,
the Green's function we used as a building blocks  should respect the vectorial as well as axial Ward identities as a constrain.
In the case of electromagnetic form factors the working expansion is known \cite{BHKRT2005,MATA200b} and here we will consider only the 
first term, which defines the so called (dressed) Relativistic Impulse Approximation (RIA).
This matrix element reads
\bea \label{form}
{\cal J}^{\mu}(p,Q)&=&eF_{\pi}(Q^2)p^{\mu}
\nn \\
&=&\frac{2N_c}{3}i e\int \frac{d^4k}{(2\pi)^4} tr\left[G^{\mu}_{EM,u}(k+Q/2,k-Q/2)
\Gamma_{\pi}(k_{r\pi_-},p+Q/2)S_d(k+p)\tilde{\Gamma}_{\pi}(k_{r\pi_+},Q/2-p)\right]+
\nn \\
&+&\frac{2N_c}{3}i e\int \frac{d^4k}{(2\pi)^4} tr\left[G^{\mu}_{EM,u}(k-Q/2,k+Q/2)
\Gamma_{\pi}(k_{r\pi_-},p+Q/2)S_d(k-p)\tilde{\Gamma}_{\pi}(k_{r\pi_+},Q/2-p)\right]
\nn \\
&+&\frac{N_c}{3}i e\int \frac{d^4k}{(2\pi)^4} tr\left[G^{\mu}_{EM,d}(k_+Q/2,k_-Q/2)
\tilde{\Gamma}_{\pi}(k_{r\pi_-},Q/2+p)S_u(k+p)\Gamma_{\pi}(k_{r\pi_-},Q/2-p)+... \right] \, ,
\nn \\
&+&\frac{N_c}{3}i e\int \frac{d^4k}{(2\pi)^4} tr\left[G^{\mu}_{EM,d}(k_+Q/2,k_-Q/2)
\tilde{\Gamma}_{\pi}(k_{r\pi_-},Q/2+p)S_u(k+p)\Gamma_{\pi}(k_{r\pi_-},Q/2-p)+... \right] \, ,
\nn \\
\eea   
where the expressions in the first (second) two lines correspond with diagrams where  the photon with momentum $Q$ couples to the up(down)-quark with electric charge $2/3e(1/3 e)$. In   Eq. \ref{form} $S_u$ stands for the up quark propagator,  Q is the photon momentum and $\Gamma_{\pi}(a,b) $ is the  pion vertex function with $a(b)$ being the relative (total) momentum of quark-antiquark pair.       
 The second line represents the triangle diagram, which  has the opposite circulation of momentum  (compared to the first one, and we also  flip the sign by taking$k \rightarrow k$) . Although we write them explicitly here, it is not difficult to show they contribute equivalently, being individually proportional to the relative momentum of the pionic pair $p$ and   pionic form factor $F$.
Thus, up to the charge prefactor, there are four identical contributions in the isospin limit, for which the propagators of light quarks are equal by definition,
 $S_u=S_d$. All propagators and vertices are dressed. For a diagrammatic representation of above see for instance \cite{YPNFS2021}.  The bare BSE vertices
 are solution of BSE with the vertex function on the right-hand  side of the BSE, being in fact identical to the BSE vertex in the approximation employed here.   
 

 The matrix $G^{\mu}_{EM}$ at each line in the Eq. (\ref{form}) is the quark-photon semi-amputated vertex defined as
 \be
 G^{\mu}_{EM,q}(k_-,k_+)=S_q(k_-)\Gamma_{EM,q}^{\mu}(k,Q)S_q(k_+);
 \label{untrun}
 \ee
 where  $ k_{\pm}=k\pm Q/2$ stands for the momenta of fermionic lines ,and  where the proper vertex  $\Gamma^{\mu}_{EM}$  is  determined by its own inhomogeneous BSE, which reads
\be \label{inBSE}
\Gamma_{EM}^{\mu}(k,P)=\gamma^{\mu}+ i \int \frac{d^4l}{(2\pi)^4} S(l_+) \Gamma_{EM}^{\mu}(l,P) S(l_-) K(l,k,P) \, ,
\ee
 where we have omitted the quark flavor $q$.  Different flavor  combinations enter various  form factors of meson, and since flavor is  not mixed by our choice of interacting  kernels $K$,  we will always mean a single quark flavor quark-photon vertex.

Thus, in order to evaluate  the form factor in Eq.  (\ref{form}), one needs to know the quark propagator $S$ , the pion Bethe-Salpeter vertex function $\Gamma_{\pi}$ as well as  the quark-photon vertex (\ref{inBSE}). In the isospin limit the propagators of $u$ and $d$  as well as the quark-photon vertices of the $u$ and $d$ quarks are identical and  
 by applying charge conjugation one can show that the second line, up to a different prefactor, which turns out to be $+1/3e$, is equal to the first one. 
The approximated set of equations for the pion  vertex  and the quark propagator we used here, were obtained in Refs. \cite{VHP2020,TPF2020} and will be  described in the following section.

\section{Light quark propagators and the pion vertices}

To get the solution for the functions $S(k)$ and $\Gamma_{\pi}(k,P)$ 
we use the simultaneous solution of DSE for the quark and BSE for the pion, and thus we  follow quite a  common practice used  in Refs.
 \cite{VHP2020,MATA2002,JIMA2001,JAMATA2003,HKRMW2005,BPT2003,dresden1,dresden2,MR1997,MT1999,EACK2008,VS2014,HGKL2017,SAJPSI,HPRG2015,HGK2015,QRS2019}.

The BSE for the vertex function $\Gamma_{\pi}$ reads  
 \bea \label{BSE}
\Gamma_{\pi}(p,P)&=&i  \int\frac{d^4k}{(2\pi)^4}\gamma_{\mu}S_q(k_+)\Gamma_{\pi}(k,P)S_q(k_-)\gamma_{\nu} [-g^{\mu\nu} V_g(q)
-C_{\Gamma}\frac{q^{\mu}q^{\nu}}{(q^2)^2}] \, ,
\eea
where the momentum $q=k-p$ and we label $C_{\Gamma}=4/3\xi g^2$, being thus identified
with the longitudinal part of the gluon propagator in a class of linear covariant gauges.
The first term should not be confused with propagator at all, albeit  
its momentum dependence of the  above kernel   was chosen to mimic the so called ladder-rainbow approximation with one gluon exchange and reads 
\bea \label{potent}
V_{g}(q)&=&\int d \om \frac{\rho_g (\om)}{q^2-\om+\ep}
\nn  \\
\rho_g(\om)&=&c_g [\delta(\omega- m_g^2)-\delta(\omega-m_L^2)] \, \, .
\eea
and it is inspired by solution of DSEs for the gluon DSE in the Landau gauge \cite{Sauli2012}.
  This model was found particularly useful 
for relatively large region of nontrivial couplings  $C_{\Gamma}$  requiring a  certain departure from popular Landau gauge 
which is our convenient strategy.  Obviously, for non-trivial $\xi$, the effective kernel $V_g$ is  gauge fixing dependent.

 We found that the presence of non trivial longitudinal modes  improves the  convergence of the solution in the  form of  the IR for the quark propagator. This IR reads
\be  \label{spectral}
S(k)=\int_0^{\infty} d x \frac{\not k \rho_v (x)+\rho_s(x)}{k^2-s+\ep} \, .
\ee
where two functions $\rho_v$ and $\rho_s$  fully characterize the quark propagator. 
  
Notably, the longitudinal part of the kernel is the only source of UV divergence in  the presented model, which was removed by dimensional renormalization.

In  Eq. (\ref{BSE}) $P$ is the total momentum of the meson satisfying $P^2=M^2$, $M=140 $ MeV for the ground state, and  the arguments in the quark propagator are $k_{\pm}=k\pm{P/2}$.  The DSE/BSE system provides precise solution 
for the quark propagator calculated in  the  gauge   $C_{\Gamma}/(4\pi)^2=0.18$  and the  kernel couplings  (\ref{potent}) $c_g/(4\pi)^2=-1.8$ and   $m_g^2/m_L^2=2/7.5$ where  $m_g$ in  physical 
units is  $m_g=556  $ MeV.


The pion  BSE vertex  function $ \Gamma_{\pi}(P,p)$ is composed from the four scalar functions   
\bea
\Gamma_{\pi}(P,p)=\gamma_5\left(\Gamma_E(P,p)+\not p \Gamma_F(P,p)+\not P \Gamma_G(P,p)+[\not p,\not P] \Gamma_H(P,p)\right) \, ,
\eea
where all of them are used to determine the pion mass, and all of them contribute to the electromagnetic form factor.
In our exploratory study presented here we simplify and use  only the first component formally.

\section{IR derived from DSEs and their use in calculation of the  function $F_{\pi}$ }

A sort of Nakanishi IRs, originally developed for scalar theories \cite{Naka1961} is 
slowly getting more use in nonperturbative   settings  of QCD \cite{VHP2020,TPF2020,YPNFS2021,SA2008,MMFP2022}, needless
to say a certain controversy on existing actual  analytical forms exists \cite{ZRC2021}.
 Independently of the detailed from of 
IRs for Green's functions in QCD and the  Standard Model, their important property  is their 
great role in performance of analytical integration in momentum space.

 IRs for Green's functions in quantum field theory play an important role since they allow analytical integration.
We  perform momentum integration  in  Eq.  (\ref{form}) analytically by using the well known formula for the Euclidean space momentum integral.
 For this purpose we  employ IRs for all   functions needed, more concretely, we use
the IR for the quark propagators in Eq. (\ref{spectral}) with the solution for $\rho_{v,s}$ as obtained for instance in the work \cite{VHP2020}.
 Motivated by the following chiral limit  Goldberger-Treiman-like identity,
\be \label{GTI}
\Gamma_E(0,p)=\frac{B(p)}{f_\pi}\, ,
\ee
where the scalar function $B$ appears in the inverse of the quark propagator:
\be
S(p)^{-1}=\not p A(p)-B(p) \, ,
\ee
hence we use a simplified version of the BS vertex, which reads
\be  \label{gold}
\Gamma_{\pi}(p,P)=\gamma_5 \frac{1}{\cal {N}}\int_0^{\infty} do \frac {\rho_B(o)}{p^2-o+\ep} \, ,
\ee
and was used with ${\cal{N}}$ being the normalization factor satisfying approximately ${\cal {N}}=f_{\pi}$
with its exact value dictated by  the canonical normalization of the BSE vertex.

 The last missing ingredient is  the  quark-antiquark-photon vertex  $ \Gamma_{EM,f}^{\mu}$ 
 for which we derive its own IR in  Appendix A.  The  version for semiamputated  vertex [Eq. \ref{untrun}] reads
 \bea 
 G_{EM}^{\mu}(p_-,p_+)&=& \sum_{i=1}^{8}  V_i^{\mu} T_i(p^2,p.Q,Q^2)
 \nn \\
 &+&\int_0^{\infty} d \om \int_{-1}^{1} dz  \frac{ \rho_v(\om) [\not{p_-} \gamma^{\mu} \not{p_+} +
\om \gamma^{\mu} ]+\rho_s(\om)[\not{p_-} \gamma^{\mu} +\gamma^{\mu} \not{p_+}]}
 {[p^2+p.Q z+Q^2/4-\omega+\ep]^2}  \,
 \label{choice}
 \eea
 where, as we show in Appendix, the second  line is in fact equivalent to the Gauge Technique Ansatz and the first line completes the entire expression 
 by adding  all transverse components independently. The eight transverse components  satisfy  the condition of transversality $V.Q=0$ and their concrete form is a matter of convention. Their convenient  representation  can be  chosen in the  following way
 \bea
V_1^{\mu}=\gamma^{\mu}_T  \, &;& V_5= p^{\mu}_T \, ;
\nn \\
V_2^{\mu}=p^{\mu}_T\not p  \, &;& V_6=[\gamma^{\mu}_T,\not p]
\nn \\
V_3^{\mu}=p^{\mu}_T\not Q  \, &;& V_7=[\gamma^{\mu}_T,\not Q] ;
\nn \\
 V_4^{\mu}= \gamma^{\mu}_T[\not Q,\not p] \, &;& V_8=p^{\mu}_T \not p \not Q \, .
 \label{baze}
\eea
 and  the associated scalar functions $T_i$ satisfies 3-dimensional integral representation
 \be
 T_i(p^2,p.Q,Q^2)= \int_0^{\infty}   d\om \int_1^{\infty} d \alpha \int_{-1}^{1} dz 
 \frac{\rho_{i,[2]}(\om,\alpha,z)}{[p^2+p.Q z+\frac{Q^2}{4}\alpha-\omega+\ep]^2} \, .
 \ee
 
 We recall that   all functions $T_i$ are for a given gauge  uniquely determined by the theory (by the solution of DSEs) through the solution for $\rho_i$.
 Also, note that  somehow arbitrary momentum-dependent prefactors used elsewhere in decomposition [Eq. (\ref{choice})] are not allowed here, 
unless they fulfill herein proposed IR.  

 There exist obviously a set of equivalent choices, depending on which part of the transverse components
 is added to the term which is fixed by gauge covariance. Other definitions of IR are possible and even the single longitudinal component 
 $\gamma_{\mu}\rightarrow \frac{Q^{\mu} \not Q}{Q^2}\ $ can be used to express the IR  
 which is fixed by  Ward identities.  We do not know yet, which choice is more advantageous from other perspectives, e.g.
 which is more suited for numerical solution. As shown in the Appendix, we have chosen  the  Gauge Technique inspired form as a tribute
  to the first nonperturbative solution of DSE for the gauge vertex appearing in the literature \cite{GT1963,GT1964,GT1977,GTII}. 
 Another  advantage is that Gauge Technique  reduces the proper vertex to the $\gamma$ matrix in the limit of vanishing 
 gauge couplings (all of them in our case).
 
   The value  $N=2$ was chosen to derive  the form of IR [Eq.  (\ref{choice}) from the DSE Eq. (\ref{inBSE})].  Since the DSE is the equation for the proper vertex, 
 the appropriate IR for this vertex is derived in the first step. Only then, it is shown the derived IR for the proper function $\Gamma^{\mu}_{EM}$
 is equivalent to the proposed IR (\ref{choice}) for the semiamputated vertex $G^{\mu}_{EM}$.


\subsection{Calculation of $F$}

\begin{figure}
\centerline{\includegraphics[width=8.6cm]{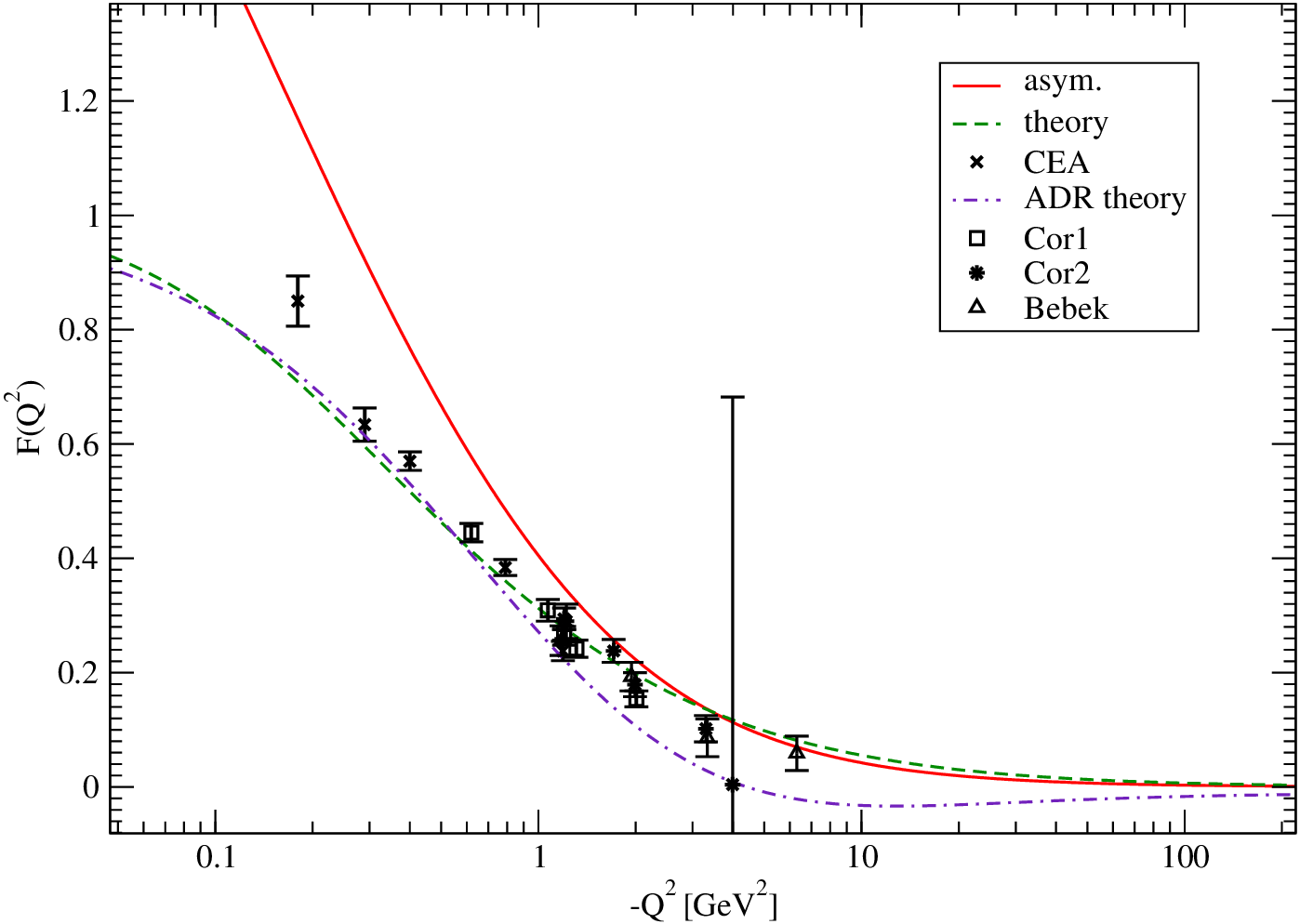}}
\caption{Calculated pion electromagnetic form factor for $Q^2<0$ and  comparison with experiment and asymptotic prediction.
The line labeled by ADR stays for evaluation based on further  approximations needed to evaluate spectral function in dispersion relation for $F$. 
For the asymptotic ( upper red line) we have chosen  the function $ F_{asym}(t)= \frac{64\pi^2f_{\pi}^2}{9\, t\,  L_a }(1+0.1 L_a^{-0.1})^2 \, ; \, L_a=ln(e+t/\Lambda^2{QCD}); \,
 \Lambda_{QCD}=250 MeV$, which obviously has a correct asymptotic $\ref{QCDPT}$.   Experimental point are: cross \cite{CEA}, square \cite{cornell1}, stars 
 \cite{cornell2} and triangle for data from \cite{bebek1978}.   }
\label{kocka2}
{\mbox{-------------------------------------------------------------------------------------}}
{\mbox{-------------------------------------------------------------------------------------}}
\end{figure}

In order to get the form factor, we use a quite primitive, albeit not easy approach, and  as we use the IRs for all vertices, we match their denominators of them by using Feynman paramaterization.
 This allows the  shifting of the loop integration momentum, and we integrate analytically in momentum space. After the momentum integration,  the resulting integral involves 9 dimensional integral 
 over the  variables  of various IRs. The integrand is highly singular for $Q^2>0$   thus being not useful in its instant form that we arrive in after  momentum integration. 
 Hence, in order to reduce the number of numerical integrations, we use further ``gauge technique approximation'', which reduces identical pairs of IR weight functions to the same number of single functions.
  Thus, for instance
\be
\int da db \rho_v(a)\rho_v(b)\rightarrow \int da \rho_v(a) \, ,
\ee
with all $b'$s replaced by $a'$s in the integral kernel. It allows further integration over auxiliary Feynman variables,  provided we are left with 
a 5-dimensional integral at the end. The appropriate derivation is relegated in the Appendix. Furthermore, since  weights of BSE vertex function
are much less known then the spectral function of the quarks we  use the integral reduction  for purpose of numerical evaluation here.
Numerical results are presented in Fig. \ref{kocka2}  for spacelike momentum, where we compare with the experiment. 
The systematic error is estimated  to be around a few percentage at  several GeV, however adjusting $F(0)=1$ is needed as we the proper renormalization 
does not lead to the correct value automatically.  

\begin{figure}
\centerline{\includegraphics[width=8.6cm]{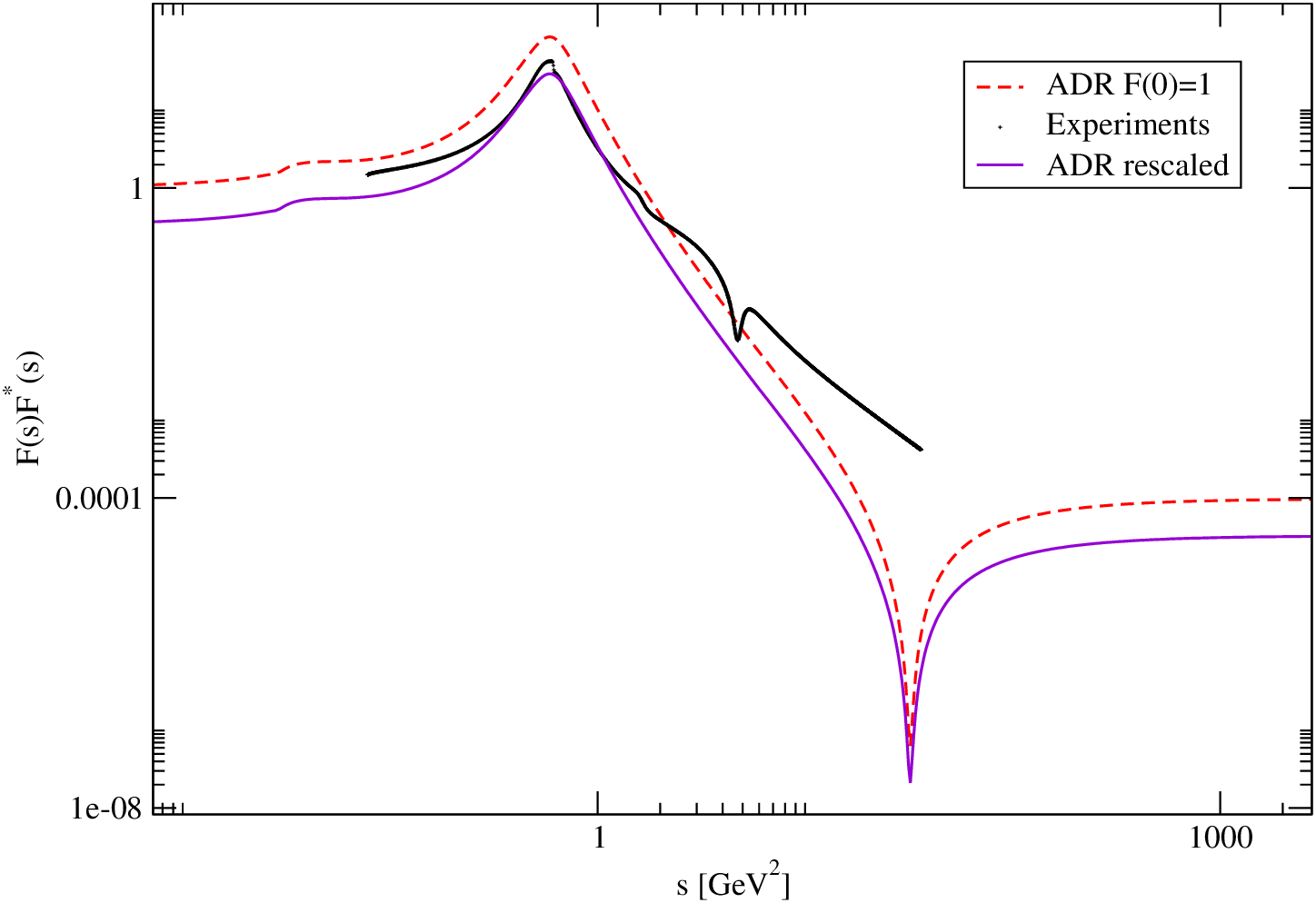}}
\caption{Calculated magnitude of the pion electromagnetic form factor for $Q^2>0$ and  comparison with experiments. The error bars are not shown, they are  within the visible size of the line and are much smaller then the deviation of presented calculations.  
The solid line is rescaled by a constant as described in the text.}
\label{kocka}
{\mbox{-------------------------------------------------------------------------------------}}
{\mbox{-------------------------------------------------------------------------------------}}
\end{figure}

Using some further approximations we derive the dispersion relation (\ref{DR}) and provide the first estimate for the resulting spectral functions of the pion electromagnetic form factor. The result is valid for low momentum and it consists of the two following terms 
\be
  \label{DR2}
  F(q^2)=\int_0^{\infty} d\omega \frac{g_1(\omega)}{q^2-\omega+\ep} 
  +q^2\int_0^{\infty} d\omega \frac{g_2(\omega)}{q^2-\omega+\ep} 
  \, ,
  \ee
where the first term could be responsible for correct normalization $F(0)=1$ if no approximation (linearization) is made. 
The functions $g_1$ and $g_2$ are given by an expression involving only a single 
integration due to the approximation employed. The result is not exact and it suffers from systematic error due to some ignored terms; however it is enough 
to show that the form factor develops the $\rho$-meson peak. In fact, the Gauge Technique approximated vertex 
is enough to get  almost the entire structure of the $\rho$ meson peak and we ignore all other transverse vertices at this stage. In order to support this statement quantitatively,  the dominant contribution 
to  $F_{\pi}(Q^2)$ has been calculated  numerically and its square is compared with world averaged
experimental data in  Fig. \ref{kocka}.  The averaged  data of the BABAR, BESS, CMD/SND and KLOE experiments \cite{I} were fitted as  described in  \cite{oser2017},
noting that there is  negligibly small experimental  error $0.5\% $ on the $\rho$-meson peak.


Within the used approximation the derivation of the desired dispersion relation (\ref{DR2})  is quite straightforward, albeit a bit lengthy and it is delegated  to  Appendix C of this work.

\begin{figure}
\centerline{\includegraphics[width=8.6cm]{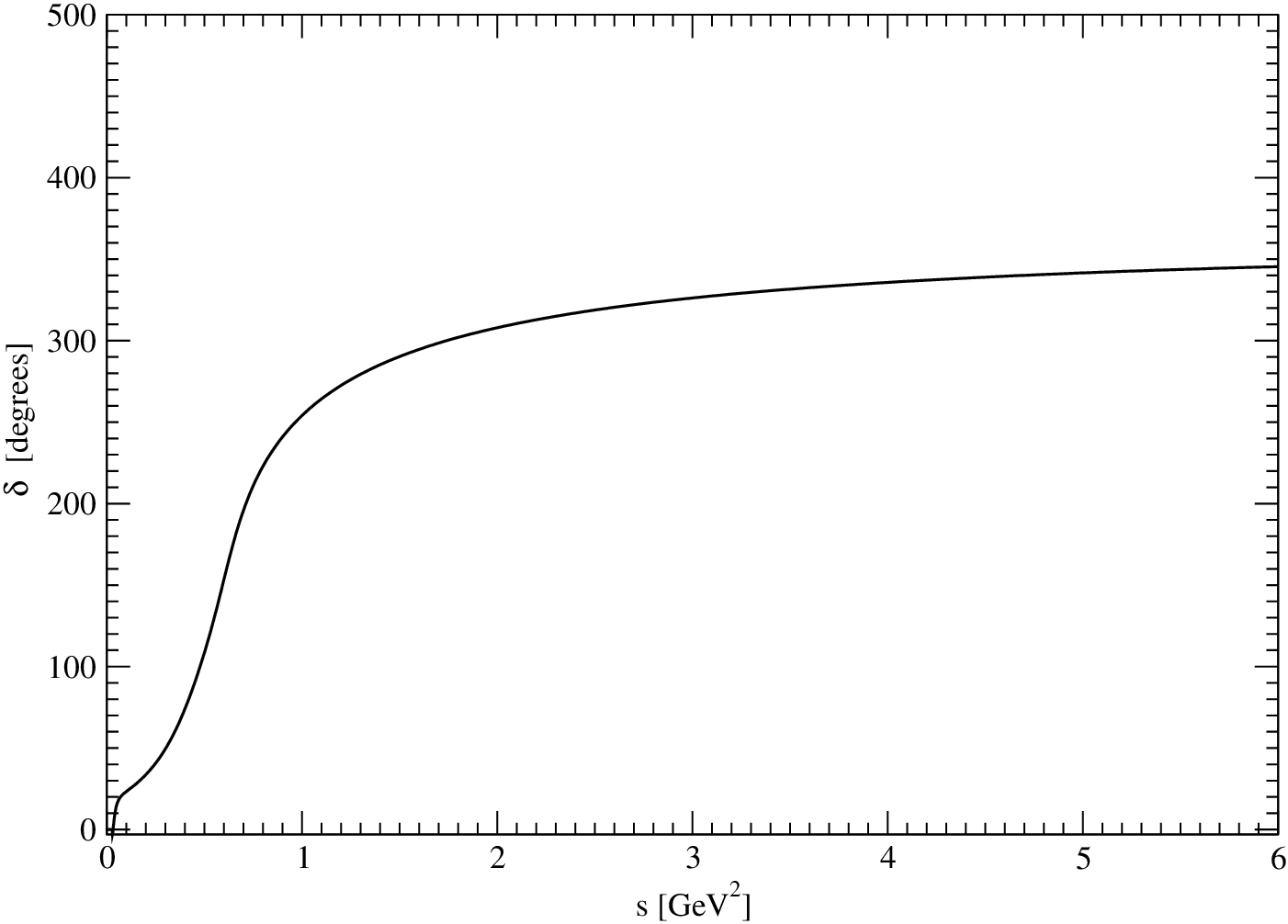}}
\caption{Phase of the pion form factor $F$ as obtained here.}
\label{kocka3}
{\mbox{-------------------------------------------------------------------------------------}}
\end{figure}

The phase $\delta$  of the form factor $F=|F|e^{i\delta}$ is shown in  Fig \ref{kocka3}.  It overestimates the phase  obtained by other methods, but still being satisfactory representative in our initial study. 
From the obtained phase we estimate that the  systematical error can be as much as $30 \%$, which is caused by linearization and other cruel approximations we made. We assume the magnitude posses the same systematics and that it 
overestimates the experimentally measured magnitude $F$, if the  condition $F(0)=1$ is imposed on the approximated form factor. 
 We lower  $F$ by a scale factor $\sqrt{3}$ for purpose of better comparison. Hence there are two  lines representing the identical result obtained by our Approximate Dispersion Relations, the dashed 
 corresponds to the standard normalization $F(0)=1$, the solid line  represents the same calculated result, but shifted down due to rescaling. 
   For higher $Q^2$ the result obtained from the dispersion relation 
 becomes untrustworthy due to our pure approximation. Obviously the first line is correct at zero momenta, while the second one reasonably approximates the peak. The difference
 is  systematic error, which is quite large in the present example.  This error suppression  is an open task and remains  a  future challenge. 
 
 Furthermore, we use the approximated DR and evaluate the form factor in the spacelike domain as well. 
 W recall that this approximation differs with the previous one and we add this result to  Fig. \ref{kocka2} for comparison.

Needless to say, the inclusion of transverse quark-photon form factors could improve the picture,and going 
beyond isospin approximation could leave some nontrivial imprints on the form factor shape.  Some part of the  systematical error could be 
due to this missing contribution, however the missing off-peak contribution is difficult to explain as due  only  to the absence of transverse quark-photon 
components.   To get rid of this uncertainty, the  developed IR in previous section could be used. We expect that the first interesting  solution for vertices will be found in the next  decade.   
 Beyond our isospin approximation, further integrations (at least two) should appear in practice
due to the necessity to use  a more sophisticated but unluckily also a more dimensional  IR \cite{SA2008,MMFP2022}
in order to evaluate the electromagnetic form factor in nonsymmetric case.

 We do not use  IRs for the transverse vertices in the part devoted to the numerical study of $F$ due to our simplified approximation. However, the revelation   
of their entire structure represents important theoretical hint for  future studies.

\subsection{Renormalization within IRs}
 
The use of the proposed IR allows the dimensional regularization to be used to regularize four-dimensional momentum integrals when they show UV divergence.

Regarding the renormalization of the vertex
the only allowed UV infinities could be associated with  $\gamma$ matrix structure, since transverse form factors have no associated terms
 in the Lagrangian of the Standard Model. 
However note, that the second and  the eighth transverse components  in the list (\ref{baze}) can, in principle,  spoil the 
renormalization properties for our (till now preferred) choice of power $N=2$ in the denominator of the IR.
  Actually, such naive UV divergences appear and it turns out, that they  cancel neither mutually nor  against
the UV term generated by Gauge Technique. 

  One possibility to get rid of UV divergences from the beginning  is that a more general $N$ can be equivalently considered. Assuming different $N$ is allowed and describes the same form factor 
\be
T_j(k^2,k.Q,Q^2)= \int_0^{\infty}   d\om \int_1^{\infty} d \alpha \int_{-1}^{1} dz 
 \frac{\rho_{j,[N_j]}(\om,\alpha,z)}
 {[k^2+k.Q z+\frac{Q^2}{4}\alpha-\omega+\ep]^{N_i}} \, ,
 \ee 
 Then Nakanishi's distributions  $\rho_{N}$ with a different  integer parameter $N$ are related through the following relation 
 \bea
 \rho_{T[N-1]}(\om,\alpha,z)&=&\frac{-1}{(N-1)}\frac{d \rho_{T[N]}(\om,\alpha,z)}{d \om}\, ;
 \nn \\
  \rho_{T[N+1]}(\om,\alpha,z)&=&-N\int_0^{\om} d o \rho_{T[N]} (o,\alpha,z)\,  .
 \eea
 where  we   assume the Nakanishi weights  vanish at boundaries.
 
Higher values of $N$ are formally allowed; however they would complicate the
future evaluation of the hadronic form factor,  so we  stay with $N=2$ here. 
To make our calculation meaningful for such a low $N$ we need to prevent the theory from unwanted UV divergences 
by another way.
For this purpose 
 one needs to impose the following sum rules:
\be
 \int_{\Gamma}   d (\om,\alpha,z)
 \rho_{2,[2]}(\om,\alpha,z)= \int_{\Gamma_3}   d (\om,\alpha,z)
 \rho_{8,[2]}(\om,\alpha,z)=0
\label{cond2}
\ee
for two weight functions  of potentially dangerous transverse form factors. 

In the  Eq. (\ref{cond2}) we  have  introduced the abbreviation for the 3 dim integration 
\be
 \int_{\Gamma_3}   d (\om,\alpha,z) f \equiv \int_0^{\infty}    d\om \int_1^{\infty} d \alpha \int_{-1}^{1} dz  f
\ee
which will be used  for the purpose of brevity.

 Actually, the combination of $T_2$ components with the gauge term of the interaction kernel then produce
 UV divergence, which is  proportional to   $\gamma^{\mu}_T$, i.e. to the first component of  the proper vertex.  Similarly
 the eighth component, which is quadratic in the relative momentum $p$ of the quark-antiquark pair (  $\simeq p^{\mu} \not p \not Q$)
 provides linear divergence in the proper vertex. In the dimensional regularization scheme it has the form:
 \be
\frac{\xi g^2}{12\pi^2} (\gamma^{\mu}\not Q-Q^{\mu})(\epsilon^{-1}+{\mbox {finite}})
 \int_{\Gamma_3}   d (\om,\alpha,z) \rho_{8,[2]}(\om,\alpha,z)
 \ee 
The derivation of entire contributions to quark-photon vertex due to the gauge interaction is delegated  shown in the Appendix.
 Quite generally, within the condition (\ref{cond2}) one makes our vertex DSE finite 
 within all transverse components properly  accounted in .

\section{Summary and Discussion}

Our results are a strong hint that there exists a consistent Integral Representation of QCD Green's functions.
If so, it is of a great interest to explore the physical predictions or within their use to calculate physical processes
already experimentally measured, but were beyond theoretical capabilities due to  timelike character of momenta in nonperturbative 
low energy strong regime of QCD.  
We have already derived Integral Representations for the quark-photon gauge vertex showing it contain  part, which is  identical with the Gauge Technique.
Within two approximations we obtained the result for the pion form factor, yet without inclusion of other transverse components of the vertex.
The spectral function of the pion electromagnetic form factor has been obtained from the applications of Integral Representation to Dyson-Schwinger
 equations for the first time. Up to the norm, the form factor agrees with the experimental data at low $Q^2$ in the both spacelike
 as well as timelike domain of momenta.  
To this point, let us mention that the  Gauge Technique  was  used in   the so called  Spectral Quark Model studies in  \cite{AB2003,MRSB2004,ABG2007},
wherein no further transverse vertices were needed to describe the broad shape of $\rho$ meson peak in calculated electromagnetic pion form factors.
 Although the Spectral Model does not solve the equations of motion for propagators, 
neither it uses the lattice predictions for this purpose, nevertheless a prognostic  feature of Spectral models 
was that the simple vertex solely dictated by the Abelian gauge invariance could be  enough for a gross description.
In this paper,  we extend the study  of \cite{AB2003} in a sense the quark propagators were calculated from the set of QCD DSEs
and within a certain ambiguity we confirm the Gauge Technique is enough to provide the gross shape of the pion form factor.

  Of course, deficiencies are due to missing 
 $\omega$ meson  and  due to the absence of isospin symmetry violating contribution.
  Further shortcomings, e.g. incorrect rate   $F_{\pi}(0)/F_{pi}(m_{\rho})$ appears due to the 
  approximations, e.g. due to the  linearization we have used at this stage.  
  Also, the phase  follows the Watson theorem very freely. In our case we get $\delta=250$ at  $1GeV$ which overestimate the values of others
 $\delta=150$. Actually, the number of numerical integrations required for the evaluation of any hadronic form factor 
 is the main weakness of proposed method. It is not the nonperturbative evaluation of building blocks: QCD vertices and propagators 
 where the calculations  is stuck, but the evaluation of form factor, where large number of entering Green's functions limit
the evaluation. Further improvement of calculation technology, e.g. avoiding a cumbersome number of auxiliary Feynman integrations till now needed
 for evaluation of hadronic form factor, is a great theoretical challenge
for future. Perhaps, a possible  generalization of old fashionable Cutkosky rules, 
seems to be a promising theoretical direction to deal with the problem more efficiently.
 
 We expect an improvement after the smooth  and more realistic version of the kernel (\ref{potent}) will be used.  More improvements can be achieved 
 when a correct weight function $\rho_{\pi}(a)$ or rather its two-dimensional form $\rho_{\pi}(a,z)$  of the pion BS vertex IR 
 \be \label{sophi}
\Gamma_{E}(p,P)=\gamma_5 \frac{1}{\cal {N}}\int_0^{\infty} d o \int_{-1}^{1} dz \frac {\rho_E(o,z,m_{\pi})}{p^2+p.P z+m_{\pi}^2/4-o+\ep} \, ,
\ee
 will be included.

 To get the desired analytical  form factor in the  Minkowski space, recall that at least quark propagator needs to satisfy a standard two body dispersion- a
  generalized K{\''a}ll{\'e}n-Lehmann representation, although the quark spectral function does need to be 
  positive definitive function. Most importantly, no other singularities but the single cut is allowed. Nontrivially,
  here we achieve this goal by our choice of the  quark-antiquark interaction kernel.

  In this respect, for many other DSEs/BSEs studies presented in the literature:
 \cite{MR1997,MT1999,MATA2002,JIMA2001,BPT2003,JAMATA2003,HKRMW2005,EACK2008,BEIVRO2011,HGK2015,HPRG2015,HGKL2017,QRS2019} , which 
 are  based on the popular version of Maris-Tandy (MT) interaction kernel introduced in \cite{STCA1990}, the proof of dispersion 
 relation could be more complicated. And at least the derivation of dispersion relation used here would invalidate at very beginning.  
 
 Recall, due to the the Gaussian kernel used in MTs, the interaction strength of MT BSE kernel blows up at timelike infinity. 
 This leads to the known  behavior: the analytical continuation of  quark propagators  exhibits infinite number
 of complex conjugated poles \cite{BPT2003,dresden1,dresden2,WI2017}.
 Such  propagators are not analytic in the domain required for the existence of the IR (\ref{spectral}) and one can repeat again, the derivation
  presented here would  invalidate from very beginning.

Our modeled DSE/BSE interacting kernel is certainly very primitive, but it includes the important ingredient- 
purely longitudinal interaction. While there should not be too much interesting physics contained in it, its numerical 
presence ensures that the Ladder-Rainbow approximation is working in the entire domain of Minkowski momentum space. 
Identifying a concrete  numerical value of  gauge parameter requires further knowledge about  other QCD vertices,
 which is out of model reach.
 However newly, in order to exhibit approximate gauge fixing independence of presented model, 
 we have  changed the gauge  fixing parameters (the entire coupling $C_{\Gamma}$).
 Thus we solve the system  numerically  in a new gauge once again, determine  a gauge dependent coupling $c_g $ and  
 in a new  gauges we  calculate the function $F$.  In all cases, only this single   parameter   was varied in order to meet pionic observable:
  the pion mass and pionic weak decay constant. The shape of the  function $F$ has been reproduced in several different gauges, showing the model 
  is actually the model of quantum gauge theory: the QCD. We plan to perform similar study within improved  setup of BSE vertices.   

\section{Acknowledgments}
I thank Ji\v{r}\'i Adam and Barbora Kub\v{e}nov\'a for
a critical reading of the manuscript and helpful discussions.

\appendix

\section{Integral Representation for quark-photon vertices }

The form of the IR for the  proper and semi-amputated photon-quark  vertex is derived in this Appendix.
 Both integral representations are related and the appropriate form are derived  as a self-consistent solution of  the DSE for the vertex (\ref{inBSE}).
The reason to keep both, the IR for proper as well as for  semi-amputated vertex  is theoretical and  practical. 
While the  DSEs are more conveniently solve in terms
of proper Greens function, the hadronic form factors are more easily evaluated in terms of semi-amputated vertices.
 
Using labeling of momentum as described in the main text, the form of Integral Representation we are going to derive for proper vertex reads: 
 \bea
 \Gamma_{EM}^{\mu}(p,Q)&=&C \gamma_{\mu}+\Gamma_{EM,L}^{\mu}(p,Q)+\Gamma_{EM,T}^{\mu}(p,Q)
 \nn\\
 \Gamma_{EM,L}^{\mu}(p,Q)&=&\sum_{i=1}^{4}  W_i^{\mu} L_{i}(p^2,p.Q,Q^2)
 \nn \\
 \Gamma_{EM,T}^{\mu}(p,Q)&=&\sum_{i=1}^{8}  V_i^{\mu} T_{i}(p^2,p.Q,Q^2)
 \nn \\
T_{i}(p^2,p.Q,Q^2)&=& \int_0^{\infty}   d\om \int_1^{\infty} d \alpha \int_{-1}^{1} dz 
 \frac{\tau_{i,[1]}(\om,\alpha,z)}{F(p,Q;\om,\alpha,z)}
 \label{choice2}
 \eea
 where 
 \be 
 F(p,Q;\om,\alpha,z)=p^2+p.Q z+\frac{Q^2}{4}\alpha-\omega+\ep 
 \ee
 for short and the individual quark momentum associated with quark legs are $p_{\pm}=p\pm Q/2$, which variable are used to label semi-amputated vertex in the main text.
 Here  $W_i$ are longitudinal matrices chosen as $1 Q^{\mu},Q^{\mu}\not p, ,Q^{\mu}\not Q$ and $ Q^{\mu}(\not p \not Q-not Q/notp) $ respectively , capital letter $V_i$ stands for  the transverse  matrix satisfying $V.Q=0$
 and $\gamma^{\mu}$ has been taken out  for calculation convenience. 
 The IR for scalar form factors $L_i$ satisfies exactly the same IR as the ones for $T_i$ , but with the distribution $\tau$ replaced by 
 its own Nakanishi weight function, say $\lambda$.
 
The bracketed index ${[1]}$  means that the first power of the denominator appearing in the last line in (\ref{choice2}) 
 has been chosen and   if not different the label  will be omitted.
  $T$ and $L$ are scalar form factors, while $V$ are for times four matrices, wherein their Dirac index are not shown for brevity, the unit i.e.  $\delta_{\alpha,\beta}$ 
 in case of the component  $V_5$ will  not be  shown as well.
 Recall,  $\tau,\lambda$ are distributions, they may involve the product of smooth functions with delta functions.

\subsection{ IR based on  DSEs and the relation with Nakanishi's PTIR}

  For pedagogical  reasons we mention the connection with the PTIR  \cite{Naka1961}  and the IR used herein.
 First of all, let us  recall here that the PTIR has been derived inductively by using perturbation theory from Feynman rules for scalar theories and 
  for various forms of PTIR   we refer to  Nakanishi's original textbook.
 
 Since the form of IR  for a Feynman diagram  is  dictated by the structure of denominators, it is very natural to assume 
   that a sort of PTIR does exist  for any renormalizable quantum field theory in 3+1 dimensions. Furthermore, it is useful to  assume (at least for a while) that the only difference 
   is that there are as many various Nakanishi weight functions as the number of independent vector/tensor matrices needed to describe a given  Feynman diagram.
   In our case of a triple fermion-gauge vertex,  there can be as much as twelve such Nakanishi weight functions $\rho_i$.
   Thus, for  each single   component the associated form factor $T$ or $L$ in $\ref{choice}$ could satisfy  the following PTIR:
 \be \label{kan}
T,L(p,Q)= \int_0^{\infty} d\omega \int d x_1 dx_2 dx_ 3  \frac{ \rho(x_1,x_2,x_3,\om)}{[p^2 x_1+p.Q x_2+Q^2 x_3-\om+\ep]} \, .
 \ee
  The polynomial structure and matrices which appear  in the numerator of  any Feynman diagrams for gauge theory vertex is dictated by Lorentz invariance and  
 are crucial for the  number of components, but  not for  the number of integral variables. Three $x-$variables are known to be  bounded as the
 Nakanishi weight function carries the delta function $\delta(1-\sum_i x_i)$, which is almost entire information we can get from analyzes of Feynman diagrams in general.  
 Obviously, by dividing by $x_1$ variable in the kernel and defining a new variables , our  proposed IR (\ref{choice}) is included in PTIR inspired from of the gauge vertex;
  however, this is far from  saying it is derivable from PTIR.

In this respect one can only say, that  the form of Nakanishi weight functions i.e., the 12  distributions of $\tau$ and  $\lambda$ can be inspected  from the perturbation theory 
expansion by studying each individual Feynman diagram in  separation. There would be a very limited benefit of doing so in strong coupling theory like QCD. 
 Therefore, our proof of the IR (\ref{choice2})  does not follow from perturbation theory, but relies on the self-consistent solution of DSE within suggested  form of  IR  implemented.
This, when embedded into the r.h.s. of the DSE for the vertex (\ref{inBSE}),  after the integration over the momentum in the Euclidean space, 
reappears on the l.h.s. of DSE again and has exactly identical form that has entered, i.e. the form of IR   (\ref{choice2}).
The set of weight functions $\rho_i$ (or equivalently  $\tau_i$ and $\lambda_i$ ) must obey certain conditions:
they satisfy  a new coupled set integro-differential  equations into which the vertex DSE (\ref{inBSE}) is transformed.
These equations do not depend on the momenta, but on the three spectral/integral variable $\omega$, $\alpha$ and $z$,
with their domain self-consistently determined by the DSEs for vertices and propagators.

For clarity, we should mention here, that the Gauge Technique  form of the vertices \cite{GT1963,GT1964,GT1977,GTII}, which was employed in calculation
of meson form factors \cite{VHP2020,TPF2020,AB2003}, represents an approximate subset of IR (\ref{choice}). 
While Gauge Technique vertices are derived Ward identities, however  they are not fully  self-consistent since they, at any known approximation of DSE, 
do generate  richer structure involving  longitudinal as well as  transverse vertices. Their entire form  is captured by three parametric IR  (\ref{choice}).

We do not know yet whether the  new -integro-differential equations for Nakanishi weights provide a unique solution, however we assume it  is the case.
 We do not even know whether functions $\rho_i$  exist at all, since  the numerical solution  is yet  out of our reach at the moment.
   However,  when keeping the solution at   hand,  as  has been already checked in 
  case of more simple truncation of DSEs system \cite{SAJPA2003,SAJHP2003}, the consistency with the  standard Euclidean formulation can be straightforwardly  inspected by the comparison.

In the next subsection of this appendix, we will illustrate the proof  on the  example of the contribution stemming from the product of the gauge technique vertex and the gauge part of the propagator as well as 
deriving  the IR (\ref{choice}) for  the particular  example of the $T_5$ component. In subsequent subsection we write down the relation between IR for proper and semiamputated vertices, which closes the proof. 
We do not provide the entire list of all contributions,  since we do not need them in our approximation. Note especially,  the conversion of  transverse pieces of $G^{\mu}_{EM}$ is a quite straightforward task and is illustrated enough
in the single  component example.

\subsection{IR for proper vertices}

The IR  has two pieces, the first  is governed by gauge covariance and the second involves all transverse components independently.
Here we show that both terms give rise to an IR for the proper vertex with the same structure of  transverse components as well as
giving rise longitudinal components in $\Gamma$. Four later longitudinal terms are in fact completely dictated by the
gauge term. We begin with transverse vertices due to the calculation simplicity. In the second part we derive IR for $\Gamma$ as it follows from 
our DSEs. 

\begin{center}{\bf Contributions to and due to the transverse vertices}\end{center}

Contributions from transverse vertices are exemplified for the most important cases. These are the ones due to the first and the fifth components, the latter 
is known to be dominant  at least in the Landau gauge . Thus in fact,  here we 
show their contributions due to other gauges. 
    
    Further, for purpose of discussion of the renormalization we also review the contribution due to the 
second as well as due to the eight component of the transverse part of the vertex.

{\bf V5:}We start with the contribution governed by the fifth component in the  Eq. (\ref{choice}), i.e. by $k_T$ where
the momentum $k$ is the relative momentum of produced quark-antiquark  pair. Contribution due to metric tensor  $\gamma\times\gamma$ part of the kernel (hence due to the
$\gamma^{\mu}_{ab}\times\gamma_{\mu;cd}/(q^2-\mu^2)$ matrices and due to the gauge part proceed similarly and we will describe the details for the first example.
The first contribution to the proper vertex in our DSE thus reads
\be
-i c_g \int\frac{d^4k}{(2\pi)^4}
\int_{\Gamma_3} d (\om,\alpha,z)
    \frac{  \rho_5(\om,\alpha,z) \gamma_{\nu} \left(k^{\mu}-\frac{Q^{\mu} k.Q}{Q^2}\right)\gamma^{\nu}}
{[F(k,Q;\om,\alpha,z)]^2 (q^2-\mu_g^2+\ep)}-... \, ,
\label{marcy}
\ee    
where three dots represent the identical  integral with the phenomenological parameter  $\mu_g$ replaced as $\mu_g\rightarrow \Lambda_g$.    
Using Feynman variable $x$ to match two denominators, making a standard square completion, shifting the integral variable and integrating 
over the momentum, we get after shome algebra, for the considered contribution:
\be
\frac{4 c_g}{(4\pi)^2}
\int_{\Gamma_3} d (\om,\alpha,z) \int_0^1 dx \frac{ \rho_5(\om,\alpha,z) p^{\mu}_T}
{p^2 +p.Qz +\frac{Q^2}{4}\frac{\alpha-z^2 x}{1-x}-\frac{\om}{1-x}-\frac{\mu^2_g}{x}}
-... \, ,
\ee    
where we have factorized prefactor  $x(1-x) $ out of  the numerator and canceled it against the same factor in the numerator 
 and where  the meaning of three dots is just  as above in  Eq. (\ref{marcy}). We do not write Dirac index and we also omit explicit writing of the Feynman
 infinitesimal term $i\epsilon$ in most   denominators for the purpose of brevity.
 
In what follows we perform the substitution  $\om\rightarrow \tilde{\om}$  and then $x \rightarrow \tilde{\alpha}$ such that
\bea
\tilde{\om}&=&\frac{\om}{1-x}+\frac{\mu^2_g}{x}
 \\
\tilde{\alpha}&=&\frac{\alpha-z^2 x}{1-x} \, ,
\label{trafo}
 \eea
 which provides the following result for the contribution (\ref{marcy})
\be
 p^{\mu}_T\frac{4 c_g}{(4\pi)^2}
\int_{-1}^{1}dz \int_1^{\infty} d\alpha \int_{\alpha}^{\infty} d\tilde{\alpha}\int_{\frac{\mu_g^2}{x}}^{\infty}
d\tilde{\om}\frac{x(1-x)^2}{\tilde{\alpha}-\alpha}
\frac{\rho_5\left[(\tilde{\om}-\frac{\mu_g^2}{x})(1-x),\alpha,z\right]}{F(p,Q;\tilde{\om},\tilde{\alpha},z)}
-... \, ,
\ee 
where the ordering of integrals is important.
 To avoid complicated explicit notation, in case the measure $dx$ is not explicitly written,   the letter $x$ will be kept 
for the  following function
\be
x=\frac{\tilde{\alpha}-\alpha}{\tilde{\alpha}-z^2} \, .
\label{xko}
\ee

Also as  follows from the inverse transformation (\ref{trafo}) the variable  $\omega_g=\omega$, which reads 
\be
\om_g=(\tilde{\om}-\frac{\mu_g^2}{x})(1-x) \, .
\label{omg}
\ee
will be kept for purpose of brevity.

In order to get the desired form of IR for proper vertex we need to change the integration ordering. Also  it is convenient to  send 
the information about integration volume into the kernel by using Heaviside step function. Performing  this correctly,
one can write for the contribution (\ref{marcy}) the resulting IR:
\bea
&&p^{\mu}_T \int_{\Gamma_3} d (\tilde{\om},\tilde{\alpha},z)    \frac{  \tau_{5a}(\tilde{\om},\tilde{\alpha},z)}{F(p,Q;\tilde{\om},\tilde{\alpha},z )}
\nn \\
 \tau_{5a}(\tilde{\om},\tilde{\alpha},z)&=&\frac{4c_g}{(4\pi)^2}\int_1^{\tilde{\alpha}} d \alpha \frac{x(1-x)^2
 \theta\left(\tilde\om-\frac{\mu_g^2}{x}\right)}{\tilde{\alpha}-\alpha}
 \rho_5[\om_g ,\alpha,z]- ...\, .
\eea
where  three dots remind us  that one should  change $\mu_g$ into the parameter $\Lambda$ appropriately, i.e, one should  introduce
a new variable
 \be
\omega_L=(\tilde{\om}-\frac{\Lambda^2}{x})(1-x) \, 
\ee
to define the new variable $\om_L$ in the function  $\rho_5[\om_L,\alpha,z]$.

Here we could stress  the difference from Nakanishi derivation   of PTIR.
Blindly following a Nakanishi's  derivation would mean the use of the variable $x$ to  give rise to our
 variable  $\omega$ (or  $\tilde{\omega}$) and we have  used slightly different strategy here.
In our approach here we avoid numerically inconvenient square roots otherwise presented in the kernel (see 
toy models without confinement \cite{SAAD2003,KUSIWI1997}).
Recall, the trick we use here, would be impossible without using the fact, that the quark propagator is entirely described by a continuous spectral function. 
In fact, this is the issue of confinement, which allows us to write the  simple equation  for IR.

{\bf V5:}     We continue with the contribution coming from the transverse vertex $V_5$ matched with 
     the gauge longitudinal interaction part of the kernel $K$. This particularly simple  contribution reads
\be
-i C_{\Gamma} \int\frac{d^4k}{(2\pi)^4}
\int_{\Gamma_3} d (\om,\alpha,z)
   \rho_5[\om,\alpha,z]   \frac{ \left(k^{\mu}-\frac{Q^{\mu} k.Q}{Q^2}\right)}
{[F(k,Q;\om,\alpha,z)]^2  q^2} \, .
\ee    
After a few  steps sketched in previous cases, this relation can be converted into the following IR:
\bea
&&p^{\mu}_T \int_{\Gamma_3} d (\tilde{\om},\tilde{\alpha},z)    \frac{  \tau_{5b}[\tilde{\om},\tilde{\alpha},z]}{F(p,Q;\tilde{\om},\tilde{\alpha},z)}
\nn \\
 \tau_{5b}[\tilde{\om},\tilde{\alpha},z]&=&\frac{C_{\Gamma}}{(4\pi)^2}\int_1^{\tilde{\alpha}} d \alpha \frac{\alpha-z^2}{(z^2-\tilde{\alpha})^2} \rho_5\left[\frac{z^2-\alpha}{z^2-\tilde{\alpha}}\om ,\alpha,z\right]\, .
\eea

Thus, for the resulting total contribution due to the fifth component one just needs to sum up
\be
\Delta \tau_{5}=\tau_{5a}+ \tau_{5b}.
\ee

Amazingly, due its simplicity, the IR  exhibits self-reproducing property: 
the contribution to the fifth component $\Delta \tau_5$  is  given by the integral over the function $\rho_5$ and no other component is generated.
However, the contribution is not complete, the contribution is not entire and other components e.g. $ p^{\mu} \not p \not Q$ component can contribute as well.
Of course , one should keep in mind,  the $\rho$ is the Nakanishi weight distribution  for the semi-amputated  vertex,
while $\tau$ is for the proper vertex.  Hence the relation between proper and semiamputated vertices  
 needs to be established. This is the subject of the second part of this Appendix.
Before that, we review other important contributions.

The transformation of contributions from terms which involve combinations of momenta is straightforward, albeit quite involved.
For the purpose of brevity we write the results in the form of fractions, which include also the second power  of $F$ in the numerator, and we also we leave   polynomial
momentum structure in the numerator  (it can be absorbed by per partes integration, which is not shown here).

{\bf V1}  In the next part  we will inspect the contribution which arises due to the first component of transverse vertex.
As the others, it is combined with the gauge part as well with the metric phenomenological interaction in the DSE for the vertex.
Explicitly written the first contribution  reads
\be
-i C_{\Gamma} \int\frac{d^4k}{(2\pi)^4}
\int_{\Gamma_3} d (\om,\alpha,z)
 \rho_1[\om,\alpha,z]   \frac{\not q (\gamma^{\mu}-\frac{Q^{\mu}\not Q}{Q^2}) \not q}
{[F(k,Q;\om,\alpha,z)]^2 (q^2)^2} \, ,
\ee    
which after repeating  similar steps as for the $V_5$ component above, leads, after some summations and trivial algebra, into the
  form:
\bea
&-&\gamma^{\mu}_T \int_{\Gamma_3} d (\tilde{\om},\tilde{\alpha},z)  \int_1^{\tilde \alpha} d\alpha
\frac{C_{\Gamma}}{(4\pi)^2} \frac{x(1-x)(2-x)}{(\tilde{\alpha}-\alpha)}
  \frac{\rho_{1}[\tilde{\om}(1-x),\alpha,z]}{F(p,Q;\tilde{\om},\tilde{\alpha},z)}
\nn \\
&-&p^{\mu}_T \int_{\Gamma_3} d(\tilde{\om},\tilde{\alpha},z)   \int_1^{\tilde \alpha} d\alpha
\frac{C_{\Gamma}}{(4\pi)^2}\frac{\rho_{1}[\tilde{\om},\alpha,z]}{[F(p,Q;\tilde{\om},\tilde{\alpha},z)]^2}
\frac{x^2(1-x)(\not Q z+2 \not p)}{(\tilde{\alpha}-\alpha)} \,  ,
\eea
noting the second power of $F$ in the second line.

Assuming the boundary condition $\rho_1(0,\alpha,z)=0$ we use per-partes integration with respect to 
$\tilde\omega$, which increases the power of $F$ by a unit.  The final,  three components IR of desired form (\ref{choice}) then read
\bea
&&\int_{\Gamma_3} d (\tilde{\om},\tilde{\alpha},z)
\frac{C_{\Gamma}}{(4\pi)^2 F(p,Q;\tilde{\om},\tilde{\alpha},z)}\int_1^{\tilde \alpha} d\alpha
\left[-\gamma^{\mu}_T \frac{x(1-x)(2-x) \rho_{1}[\tilde{\om}(1-x),\alpha,z]}{(\tilde{\alpha}-\alpha)}\right. 
\nn \\
 &-&\left.p_T^{\mu} \frac{x^2(1-x)\frac{d}{d\tilde \om}\rho_1(\tilde{\om}(1-x),\alpha,z)] (\not Q z+2 \not p)}{(\tilde{\alpha}-\alpha)}\right] \,  .
\eea

{\bf V1 g:}To calculate the  contribution due to the interaction kernel with the metric tensor is more simple. 
The appropriate contribution reads:
\be
-i C_{\Gamma} \int\frac{d^4k}{(2\pi)^4}
\int_{\Gamma_3} d (\om,\alpha,z)
 \rho_1[\om,\alpha,z]   \frac{ \gamma_{\beta} (\gamma^{\mu}-\frac{Q^{\mu}\not Q}{Q^2}) \gamma^{\beta}}
{[F(k,Q;\om,\alpha,z)]^2 (q^2-\mu_g^2)}-... \, ,
\label{bar-borka}
\ee    
Repeating  basically same steps, which were used to  
 transform the $T_5$ contribution, 
one gets for (\ref{bar-borka}) the  result
\be
-\gamma_T^{\mu}\int_{\Gamma_3} \frac{d (\tilde{\om},\tilde{\alpha},z)}{F(p,Q;\tilde{\om},\tilde{\alpha},z)}
\frac{C_{\Gamma}}{(4\pi)^2}\int_1^{\tilde \alpha} d\alpha\frac{2x(1-x)\rho_1[\om_g,\alpha,z]
\theta\left(\tilde{\om}-\frac{\mu_g^2}{x}\right)}{\tilde \alpha-\alpha} \, ,
\ee
 where since the substitution (\ref{trafo}) was made at the end of the transformation [ $x$ stands for the fraction (\ref{xko}) 
 and $\omega_g$ is defined by  Eq. (\ref{omg})].

{\bf V2} For contribution to proper quark-photon vertex due the second transverse  component (i.e. $k^{\mu}_T\not k$) and due to  the gauge interaction kernel takes the form 
\be
-i C_{\Gamma} \int\frac{d^4k}{(2\pi)^4}
\int_{\Gamma_3} d (\om,\alpha,z)
 \rho_2[\om,\alpha,z]   \frac{ \not q \left(k^{\mu}-\frac{Q^{\mu} k.Q}{Q^2}\right)\not k \not q}
{[F(k,Q;\om,\alpha,z)]^2 (q^2)^2} \, ,
\ee    
can be readily transformed into the following form
\bea
&&\gamma_T^{\mu}\frac{C_{\Gamma}}{4(4\pi)^2}(1/\epsilon_d-\gamma_E)\int_{\Gamma_3} d (\om,\alpha,z) \rho_2[\om ,\alpha,z]
\nn \\
&+& \gamma_T^{\mu}\frac{C_{\Gamma}}{2(4\pi)^2}\int_{\Gamma_3} d (\tilde{\om},\tilde{\alpha},z) 
\int_1^{\tilde{\alpha}} d\alpha \frac{x^2(1-x)}{\tilde\alpha-\alpha}
\frac{\left(R_2[\om(1-x) ,\alpha,z]-(1-x)(Q.p z+2 p^2)\rho_2[{\tilde{\om}}(1-x) ,\alpha,z]\right)}
{F(p,Q;{\tilde\om},{\tilde\alpha},z) }
\nn \\
&-&p_T^{\mu}\not p\frac{C_{\Gamma}}{(4\pi)^2}\int_{\Gamma_3} d (\tilde{\om},\tilde{\alpha},z) 
\int_1^{\tilde{\alpha}} d\alpha \frac{x(1-x)^2(1+x)}{\tilde\alpha-\alpha}
\frac{\rho_2[\om(1-x) ,\alpha,z]}
{F(p,Q;\tilde{\om},\tilde{\alpha},z)}
\nn \\
&+&2 p_T^{\mu}\frac{C_{\Gamma}}{(4\pi)^2}\int_{\Gamma_3} d (\tilde{\om},\tilde{\alpha},z) 
\int_1^{\tilde{\alpha}} d\alpha\frac{x^2(1-x)^2}{\tilde\alpha-\alpha}
\frac{\rho_2[\om(1-x) ,\alpha,z](-Q.p z/2-Q^2\tilde{\alpha}/4+\tilde{\om})(\not Q z/2+\not p)}
{[F(p,Q;{\tilde\om},{\tilde \alpha},z)]^2} \, .
\eea
We plan to publish the details of derivation and further useful relations in a separate Supplementar Material.

 The Euler constant
$\gamma_E$, which arises in the equation above, is due to the standard dimensional regularization and should not be confused with projected gamma matrices. 
As we can see the second component of $T$ gives rise to nontrivial contributions to the four different components, including the second component itself.
The elimination of momentum from the numerator and adjusting the power of $F$ to the desired value $N_2$ is matter of standard practice.
Furthermore, let us mention that due to the $z$ dependence, many terms turn out to be zero and the equation above is already in form suited for the first calculation.
 Note also, there is a single term proportional to $\gamma_T$, which is divergent in the limit $ d\rightarrow 4; (\epsilon_d\rightarrow 0)$.


\begin{center}{\bf Contributions due to the Gauge Technique vertex}\end{center}

In the following part we will inspect   the  proper vertex contribution due to the  Gauge Technique IR.
As a part of the proof,  we will show that the  Gauge Technique is equivalent to the proposed IR for semiamputated vertex.
 Then we will calculate its contribution to proper vertex for its combination with  the gauge , i.e. purely longitudinal part of the  interaction kernel.
  As  is clear fromthe  previous part devoted to the transverse components, deriving the IR
 due to the interaction with a metric tensor  is a matter of practice, where  several  well-controlled changes
 in derivation cannot violate the resulting functional form of the IR.  

 Considering the DSE  with aforementioned inputs on the rhs of the  DSE (\ref{inBSE}) means to 
 evaluate the following contribution 
\be \label{gterm}
-i C_{\Gamma}\int\frac{d^4k}{(2\pi)^4} \not q G^{\mu}_{GT}(k_-,k_+)\frac{\not q}{(q^2)^2}
\ee
where again we label $C_{\Gamma}=e_q g^2 \xi T_a^2$ and the momentum associated with the internal gluon line is $q=p-k$. 

In the first step, we will show the Gauge Technique vertex, in its  conventional form: 
\be 
G_{GT}^{\mu}(k_-,k_+)=
\int_{-\infty}^{\infty} d x \frac{ \rho(x) \gamma^{\mu}}
{[\not{k_-}-x+\ep][\not{k_+}-x+\ep]}
\ee
(see some of the papers \cite{GT1963,GT1964,GT1977,GTII}) is identical to the second line of IR for semiampuated vertex (\ref{choice}).
As we prefer to work with two quark propagator spectral function  $\rho_v$ and $\rho_s$ we rewrite the above expression into less familiar form
\be \label{normal}
G_{GT}^{\mu}(k_-,k_+)=
\int_{0}^{\infty} d a \frac{ \rho_v(a) [\not{k_-} \gamma^{\mu} \not{k_+} +
a \gamma^{\mu} ] 
+\rho_s(a)[\not{k_-} \gamma^{\mu} +\gamma^{\mu} \not{k_+}]}
{(k_-^2 -a+\ep)(k_+^2-a+\ep)} \, ,
\ee
where two  functions $\rho_v$ and $\rho_s$ are defined on $R^+$ and they are  related with the  single function $\rho$ 
in the following manner
\bea \label{shit1}
\rho_v(a)&=&\frac{\rho(\sqrt{a})+\rho(-\sqrt{a})}{2\sqrt{a}}
 \nn \\
\rho_s(a)&=&\frac{\rho(\sqrt{a})-\rho(-\sqrt{a})}{2} \, .
\eea    
The advantage of our choice is that  the function on lhs. takes a nontrivial value at the positive real axis,
 which simplifies many  manipulations we will perform and show in this appendix. 
In addition we use the following identity
\be
\frac{1}{k_-^2 -a}\frac{1}{k_+^2 -a}=\int_{-1}^{+1} d z \frac{1}{[k^2+k.Q z+\frac{Q^2}{4}-a]^2}
\ee
in order to match the denominators in (\ref{normal}), getting thus desired form 
that corresponds to  the second line of the entire IR (\ref{choice}).

In this way we have proved that the Gauge Technique is a part of proposed IR in addition to converting $ G_{GT}$
into the form suited for  evaluation. 
 Substituting Eq. (\ref{normal}) into the formula (\ref{gterm}) we get at this stage
\bea \label{guterm}
\sum_{v=a,b,c,d}\Gamma_{v}^{\mu}(k,P)&=&-i C_{\Gamma} \sum_{v=a,b,c,d}  \int\frac{d^4k}{(2\pi)^4} \int_0^{\infty} da \int_{-1}^{+1} d z
\frac{N^{\mu}_v(k,p,Q;a)}{[k^2+k.Q z+\frac{Q^2}{4}-a]^2 (q^2)^2}
\\
N_a^{\mu}&=& \rho_v(a)\not q(\not k-\not Q/2)\gamma^{\mu}(\not k+\not Q/2)\not q  \, ;\, \,
N_b^{\mu}=  \rho_v(a)\not q \gamma^{\mu} \not q \, ,
 \\
N_c^{\mu}&=&2 \rho_s(a) k^{\mu}q^2\, ;
\, \, N_d^{\mu}= \not q [\gamma^{\mu},\not Q]\not q \,.
\eea

From this point, up to the different matrix structure of the numerator, the treatment of the rest is easy as in the case of previous study of transverse contribution.
 Of course, the IR is two instead of three dimensional, thus aside from a completely continuous part, one can expect the delta function when one uses
three dimensional write up $ (\delta(\alpha-1))$.  Nevertheless, even so, the IR for the proper function turns to  be3-dimensional.

To derive the IR we will use the  variable $y$ to match the result with $q^2$ in the denominator, which leads to the following result :
\be
\Gamma_v^{\mu}(k,P)= 
-i  \int\frac{d^4k}{(2\pi)^4}\int_0^{\infty} \int_{-1}^{1} dz
\int_0^1 dy   \frac{3 C_{\Gamma} y(1-y) N_v^{\mu}}{[\left(k+\frac{Q}{2}zy-p(1-y)\right)^2-(\frac{Q}{2}zy-p(1-y))^2+p^2(1-y)+\frac{Q^2}{4}y-ay]^4} \, .
\ee
for each term defined in the Eq. (\ref{guterm}).

The rest of transformation is quite universal and we will not repeat it for all terms individually, but we illustrate it for two scalar cases.
The first is the integral where we replace the function $N_v^{\mu}$ by $N_1= f(a)$ where $f(a)$ stands for some continuous functions. Let us label such an
auxiliary scalar  vertex by the  index $1$. Integrating over the momentum we directly get
\be \label{ganul}
\Gamma_1(k,P)= \frac{C_{\Gamma}}{(4\pi)^2}\int_{-1}^{1} dz \int_0^{\infty} da f(a)
\int_0^1 dy   \frac{[y(1-y)]^{-1}}{[p^2+Q.p z+\frac{Q^2}{4}\frac{1-z^2 y}{1-y}-\frac{a}{1-y}]^2}
\ee
For completeness, we repeat the entire transformation here .

Let us perform the substitution $y\rightarrow \alpha$ such that 
\be \label{cudr}
\alpha=\frac{1-z^2y}{1-y} \, \, , \, \,  y=\frac{\alpha-1}{\alpha-z^2}.
\ee
And then the  second substitution $a\rightarrow \omega$ such that
\be
\om=\frac{a}{1-y} \, ,
\label{chcip}
\ee
obtaining thus for (\ref{ganul}) the following expression
\be
\Gamma_1(k,P)= \frac{C_{\Gamma}}{(4\pi)^2}\int_{-1}^{1} dz 
 \int_1^{\infty} d\alpha
 \int_0^{\infty} d\omega \frac{\frac{1-z^2}{(1-\alpha)(z^2-\alpha)} f\left[\omega\frac{1-z^2}{\alpha-z^2}\right]}
 {[p^2+Q.p z+\frac{Q^2}{4}\alpha-\omega+\ep]^2} \, .
\ee

As in the previous part, here the variable $y$, if used without integral measure in any expression will be kept even after the substitutions are performed 
for the purpose of brevity. Its meaning will be unique throughout this paper and always given by the second Eq. in (\ref{cudr}).

The scalar function $\Gamma_1$ is not yet in desired form  and for this purpose we perform per-partes integration with respect to the variable $\omega$.
Doing this  we can write
\be 
\Gamma_1(k,P)=\int_{\Gamma_3}d (\omega,\alpha,z)
\frac{\frac{C_{\Gamma}(1-z^2)}{(4\pi)^2(1-\alpha)(\alpha-z^2)}\frac{d}{d \om} f\left[\omega\frac{1-z^2}{\alpha-z^2}\right]}
 { F(p,Q;\om,\alpha,z)} \, ,
 \label{scal2}
\ee
 where we have assumed the function $f$ is vanishing at boundaries. Recall, within a numerical accuracy, this is certainly true for  the quark spectral function and we will repeatably exploit the fact that   $(\rho_{v,s}(0)=\rho_{v,s}(\infty)=0)$.
 
 The assiduous reader can note that there is an infrared log divergence involved in the $\alpha$ integral in Eq. (\ref{scal2}).
 These are standard IF divergences owned to massless gauge boson modes and actually similar divergences appear in  $\Gamma^{\mu}$ 
 and make associated form factor procedure dependent. If appears numerically, it could 
  be used to  cancel against  similar divergences due to the emission of soft real photons in the physical cross section.
   This fact, however, does not bother us yet, since we are  not going to  solve  the DSEs system in this paper. 

Such IF divergence does not appear for  a less divergent kernel. Therefore, by replacing  for instance  $1/q^4$ kernel in the Eq. $(\ref{ganul})$  with $1/q^2$ one can get
the IR in the following regular form:
\be 
\int_{\Gamma_3}d (\omega,\alpha,z)
\frac{\frac{C_{\Gamma}}{(4\pi)^2} f\left[\omega\frac{1-z^2}{\alpha-z^2}\right]}
 { F(p,Q;\om,\alpha,z)} \, .
\ee

{\bf C:}   
   
 Repeating  the game for our vertex (\ref{gterm}) is relatively straightforward. The only complication is that one is faced to a larger number of  
 momentum integrations over various  tensors.  
  Here we start with the simplest case, say $N_c^{\mu}$ term in the  Eq. (\ref{guterm}).
Taking  changes into account, one gets two vector contributions, the first  contributes to  the transverse  component $V_5$ (and by the same amount to the longitudinal counter-partner) and the second is purely longitudinal. Explicitly it reads  
 \be
 \Gamma_c^{\mu}(p,Q)= \frac{C_{\Gamma}}{(4\pi)^2}\int_0^{\infty} d \omega \int_1^{\infty} d \alpha  \int_{-1}^{1} dz  
  \frac{(1-z^2)}{(\alpha-z^2)^3}\rho_s\left[\omega\frac{1-z^2}{\alpha-z^2}\right]
  \frac{\left[2p^{\mu} (1-z^2)+Q^{\mu} z(1-\alpha)\right]}{F(p,Q;\om,\alpha,z)} \, .
 \ee

{\bf A}:
 
The conversion of  $\Gamma_a$ is technically the most  demanding, -not only this piece  involve UV divergence, but   a  double per-partes is needed to convert this part into the desired IR. 
Hence we will comment on some steps in more detail.

Here the UV divergent terms,
 which stem from the first terms of   numerator expansions
\bea
\not q \not Q\gamma^{\mu} \not Q \not q&=&\gamma^{\mu} Q^2 q^2+4 q^{\mu} Q.q \not q-2Q^{\mu}q^2\not Q-2q^{\mu}Q^2\not Q
\nn \\
\not q \not k\gamma^{\mu} \not k \not q&=&\gamma^{\mu} k^2 q^2+...
\eea
will be concerned here. We will not list all  IRs stemming from other contributions, which are relatively easy to evaluate;
 we will publish them when an actual numerical solution is available. 

In order to see how  individual terms arise during the derivation, we will write down a few intermediate steps.
Summing the first terms in expansions above, then we get after Feynman paramatrization (i.e. before the transformation (\ref{chcip}))  the following 
result
\be
-i C_{\Gamma}\int da \rho_v(a) \int_0^1 dx \int_{-1}^{1}dz 
\int\frac{d^4k}{(2\pi)^4} \frac{\gamma^{\mu}\left[k^2-\frac{Q^2}{4}\right] x\Gamma(3)}
{[{\tilde k}^{2}+p^2(1-x)x+\frac{Q^2}{4}x(1-z^2 x)+p.Q zx (1-x)-ax ]^3} \, ,
\ee
where $\tilde k=k+QZx/2-p(1-x) $ and we omit some prefactors for the purpose of brevity.

We will use the dimensional regularization; thus we label $\epsilon^{-1}=4-d$ as  the divergent constant in 4 dimensions.  After the usual shift the term proportional to
 ${\tilde k}^2$  gives 
\be
 \int_0^1 dx \frac{2\gamma^{\mu}x}{(4\pi)^2}\left[-\frac{2}{\epsilon}-\gamma_E+ \ln(1-x)x +\ln{F(p,Q;\frac{a}{1-x},\frac{1-z^2x}{1-x},z)}\right]\, ,
 \ee
where we omit some unimportant prefactors. 

After the substitution (\ref{chcip}) (with $x$ instead of $y$)  we get the following entire expression:
\bea
\Gamma_c^{\mu}(p,Q)&=& \gamma^{\mu} Const.+\gamma^{\mu}
\frac{C_{\Gamma}}{(4\pi)^2} \int_{\Gamma_3} d(\omega,\alpha,z) \frac{2y(1-y)(1-z^2)}{(z^2-\alpha)^2}\rho_v[\om(1-y)] \ln{[F(p,Q;\om,\alpha,z)]}    
\nn \\
&+&\gamma^{\mu}\frac{C_{\Gamma}}{(4\pi)^2} \int_{\Gamma_3} d(\omega ,\alpha,z)  
\frac{(1-z^2)}{(z^2-\alpha)^2} \rho_v[\omega(1-y)]
 \frac{\frac{Q^2}{4}(z^2 y^2-1)+p^2(1-y)y-Q.p z (1-y)y}{F(p,Q;\om,\alpha,z)} \, ,
\label{linhasec}
\eea
where in order to avoid cluttering notation, we remind here that the letter $y$ is simply (\ref{cudr})(since $x$  is reserved for a different  function
in our notational convention).
In this process, a constant term (UV divergent) $\gamma^{\mu}Const$  appears, into which we also  sent  some constant pieces,
 which have been generated during the derivation.  It should be taken  in mind that the entire i.e.  the finite as well infinite part of the vertex 
 could be consistent with the renormalization of the quark DSE due to the WTI.

 To transform the first line to the desired IR we 
use per-partes integration with respect to the variable $\omega$. For this purpose we use the following expression for the primitive function in the numerator
\be
R_v[\om,\alpha,z]=\int_{0}^{\omega} du \rho_v[u(1-y)] \, .
\ee
Further, irrespective of  the value of  boundary term, we sent it into the constant term. The remaining of the first line then reads
\be
 \gamma_{\mu} \left[Const.+\int_{\Gamma_3} d(\omega ,\alpha,z)  \frac {\rho_{log}(\om,\alpha,z)}{F(p,Q;\om,\alpha,z)}\right]\, ,
 \label{primo}
 \ee
 where the  contribution to the vertex weight function is
\be
\rho_{log}(\om,\alpha,z)=2 C_{\Gamma} \frac{y(1-y)(1-z^2)}{(\alpha-z^2)^2}R_v[\om,\alpha,z]  \, .
\ee
 
 To convert the second line in Eq. (\ref{linhasec}) one can divide term with $p^2$ as the first step. Then the term in the numerator, which is proportional to the variable $Q^2$, can be treated by per-partes integration to cancel it with the prize we get $ln(J)$ instead $J^{-1}$. In order to get  $J$ back in the denominator one 
 can integrate per-partes, but now with respect to the variable $\omega$. The terms involving scalar product $p.Q$ in the numerator can be treated analogously, but instead of the variable $\alpha$ one needs to use the variable $z$. The single resting term  has already been derived in this form of IR.
 The entire result for the second line is then given by (\ref{primo}) where instead of $\rho_{\log}$ we have the following function
\bea 
&&C_{\Gamma}  \frac{d}{d z} 
\left[\frac{(1-z^2)}{(z^2-\alpha)^2} z(1-y)(2y-1) R_v[\om,\alpha,z]\right]  
\nn \\
&+&C_{\Gamma}  \frac{d}{d\alpha} 
\left[\frac{(1-z^2)}{(z^2-\alpha)^2}[1-z^2y^2+\alpha (1-y)^2] R_v[\om,\alpha,z]  \right]
 \nn \\
&-&C_{\Gamma} \omega \frac{(1-z^2)}{(z^2-\alpha)^2}(1-y)^2 \rho_v[\omega(1-y)] \, .
\eea

{\bf D:}
 
 Repeating  the game for the last   numerator in (\ref{guterm}) gives us
 
 \bea \label{janica}
 \Gamma_{d}^{\mu}(p,P)&=&
 \int_{-1}^{1} dz 
 \int_1^{\infty} d\alpha
 \int_0^{\infty} d\omega \frac{C_{\Gamma}}{(4\pi)^2}\frac{(1-\alpha)(1-z^2)}{(z^2-\alpha)^3} \frac{M^{\mu} \rho_s\left(\omega\frac{1-z^2}{\alpha-z^2}\right)}
 {F(p,Q;\om,\alpha,z)^2} \, 
 \nn \\
 M^{\mu}&=&(\not p+\not Q/2)[\gamma_{\mu},\not Q] (\not p+\not Q/2) 
 \eea  
 where we do not write Dirac index for brevity.
 
 Using the following identity
 \bea
 M^{\mu}&=&p^2[\gamma^{\mu},\not Q]+2 p^{\mu}[\not p,\not Q]+2Q.p [\not p,\gamma^{\mu}]
 \nn \\
 &-&\frac{z^2}{4} Q^2[\gamma^{\mu},\not Q]
 \nn \\
 &+&z\left(Q^{\mu}[\not Q,\not p]+Q^2[\not p,\gamma^{\mu}]\right) \, ,
 \eea
 one can immediately recognize   various components of the quark-photon vertex, and thus for instance the last line
 when implemented in (\ref{janica})  gives $\simeq z(-1)Q^2 V_6$ in the numerator. The last step to get desired IR is the per-partes contribution with respect
 to the variable  
 $\alpha$, which lower the power of $F$  in  the denominator and cancels out unwanted presence of   $Q^2$ in the numerator.
 To convert other terms of $M$  into IR is matter of simple algebra andthe  repeated use of per-partes integration.
 The result, together with the numerical solutions, will be published in the future.   
 

\subsection{Integral representation for a semiamputated vertex}

Using an accepted form of the semiamputated vertex we have shown the proper vertex $\Gamma^{\mu}$ satisfies
the integral representation, which up to the power of the denominator has the identical form as assumed form of the semi-amputated vertex itself.
What remains is to show that the IR for SAV is consistent with the obtained IR for proper vertex from DSE solution.

The first power of $F$ in the denominator of the IR  is   preferable choice for next purpose, which choice simplifies soem parts of calculation.
However remind, $N=2$ was a preffered choice in preceding sections. Here we thus should note, that these two weight functions are simply related.

THus we are going to find relation between the IR of left and right side of the following definition: 
\be
G^{\mu}_{EM}(k^+,k^-)=S(k^-) \Gamma_{EM,T}^{\mu}(k,Q) S(k^+)
\ee
with all functions on the rhs. expressed through their own IR.
Plugging  this IR for the proper vertex, which we recall here as 
\bea
\Gamma_{EM,T}^{\mu}(k,Q)&=&\sum_{i=1}^{8}  V_i^{\mu} T_{\Gamma,i}(k^2,k.Q,Q^2)
 \nn \\
T_{\Gamma,i}(k^2,k.Q,Q^2)&=& \int_0^{\infty}   d\om_{\Gamma} \int_0^{\infty} d \alpha \int_{-1}^{1} dz_{\Gamma} 
 \frac{\tau_i(\om,\alpha,z)}
 {[F(k,Q;\omega_{\Gamma},\alpha,z)]} 
 \eea
 together with spectral representations for the quark propagators $S$
into the rhs of the SAV definition:
 \be
G^{\mu}_{EM}(k^+,k^-)=S(k^-) \Gamma_{EM,T}^{\mu}(k,Q) S(k^+)
\ee
we are  prepared to convert the resulting expression
\be  
\label{SAV}
\int_0^{\infty} da db d\om d\alpha \int_{-1}^{1} dz_{\Gamma}\frac{{\not k}_-\rho_v(a)+\rho_s(a)}{(k_-^2-a+\ep)}\frac{\tau_T^i(\om_{\Gamma},\alpha,z_{\Gamma}) V_i^{\mu}}{(k^2+k.Q z_{\Gamma}+\frac{Q^2}{4}\alpha-\om_{\Gamma}+\ep)}
\frac{(\not k_+\rho_v(b)+\rho_s(b))}{(k_+^2-b+\ep)}
\ee
into the form of suggested IR (\ref{choice}) for the lhs of the definition of the SAV.

Let us first briefly describe the core of the proof. As a first step,  we  commute all $V$'s from the middle position into the front, and  use the Feynman rules for denominators to match the propagators $S$  and the proper vertex together.
This gives us  the IR with some additional presence of the  scalar product of momenta in the numerator.
If a given term in the numerator belongs to $T_i$, we need only adjust a proper denominator $N=2$. If there is additional momentum dependence in the  prefactor, we use the per-partes integration
  to remove it with  simultaneous change of power of   the numerator. At the end, we adjust the power of the denominator to $N=2$ by per-partes integration with respect to newly defined  variable  $\omega_{new}$.
 From all of $V$ we choose the Dirac $\gamma_T$ only, the other terms  proceed similarly. In what follows we will not write the Feynman $\ep$; its presence is assumed implicitly.  

 Using the  formula
 \bea
\frac{1}{k_-^2 -a}\frac{1}{k_+^2 -b}&=&\int_1^{-1} d z_G \frac{1}{[k^2+k.Q z_G+\frac{Q^2}{4}-\omega_G]^2}
\label{doubprop}
\\
\omega_G&\equiv&\frac{a}{2}(1-z_G)+\frac{b}{2}(1+z_G)
\nn
\eea
and further matching with the denominator of the proper vertex in (\ref{SAV})
\bea
 &&\frac{1}{[k^2+k.Q z_G+\frac{Q^2}{4}-\omega_G]^2}\frac{1}{[k^2+k.Q z_{\Gamma}+\frac{Q^2}{4}\alpha-\omega_{\Gamma}]}
\nn \\
&=& \int_0^1 dx \frac{2x}{[k^2+k.Q (z_G x+ z_{\Gamma}(1-x)+\frac{Q^2}{4}(x+\alpha(1-x))-\omega_G x- \omega_{\Gamma}(1-x)+\ep]^3}
\eea
one can write the result
\bea
G^{\mu}_{EM}(k,Q)&=&\int_0^{\infty} d\alpha  da db \int_0^1 dx \int_{\noindent -1}^{1}  d z_{\Gamma} d z_G   \frac{-2x [\rho_v(a)\rho_v(b) \gamma^{\mu}(Q^2/4-k^2)+R^{\mu}]}
{[k^2+k.Q (z_G x+ z_{\Gamma}(1-x)+\frac{Q^2}{4}(x+\alpha(1-x))-\omega_G x- \omega_{\Gamma}(1-x)]^3}
\nn \\ 
R^{\mu}&=&(2k^{\mu}+Q^{\mu})(\not k-\not Q/2)+\gamma^{\mu}/2[\not k, \not Q] +\rho_s(a)\rho_s(b)\gamma^{\mu}
+\rho_v(a)\rho_s(b)(2k^{\mu}+Q^{\mu})
\nn \\
&-&\rho_v(a)\rho_s(b)\gamma^{\mu})(\not k+\not Q/2)+\rho_s(a)\rho_v(b)\gamma^{\mu})(\not k-\not Q/2)\, .
\eea

As a next step we perform the following substitutions
\bea
\tilde{\alpha}&=&x+\alpha(1-x)
\nn \\
\tilde{z}&=&z_G x+z_{\Gamma}(1-x)
\nn \\
\tilde{\om}&=&\om_G x+\omega_{\Gamma}(1-x)
\label{forx}
\eea
such that $\alpha \rightarrow \tilde\alpha$ , $z_G\rightarrow \tilde{z}$ and $x \rightarrow \tilde{\omega}$, thus we get the wanted form of the denominator. 
Doing this explicitly, we can write for the Eq. (\ref{SAV})
\bea 
G^{\mu}_{EM}(k,Q)&=&\int_0^{\infty} d \tilde{\omega} d\tilde{\alpha}  \int_{-1}^{1} \ d \tilde{z}
 \frac{I}{k^2+k.Q\tilde{z}+\frac{Q^2}{4}\tilde{\alpha}-\tilde{\om}}
 \label{fester}
 \\
 I&=&\int_0^{\infty} da db d\om_{\Gamma} \int_{-1}^{1} dz_{\Gamma}  
 \frac{-2x \prod \theta [\rho_v(a)\rho_v(b) \gamma^{\mu}(Q^2/4-k^2)+R^{\mu}]}{(1-x)[\frac{a}{2}(1+z_{\Gamma})+\frac{b}{2}(1-z_{\Gamma}-\om_{\Gamma}]}
\nn
\eea
where $x$ is the solution of Eqs (\ref{forx}) ,  i.e., it is a function $x(a,b,z_{\Gamma},\tilde{z}, \tilde{\alpha},\omega_{\Gamma})$. We label$ \prod \theta$ the product of step Heaviside functions
which define the integration domain and straightforwardly stem from the transformation (\ref{forx}). They ensure that the numerator is zero in the boundaries of three integrals appearing in  (\ref{fester}). 
To get the form of IR with desired power of the numerator, one only needs to employ pert partes integrations.

  \section{Evaluating pion form factor within the gauge technique approximation}

 The  function $F(Q^2)$ due to the gauge technique  vertex in two approximations  is derived in this appendix. Both  are based on the gauge technique -like linearization, which reduces the number of numerical integrations.
  Further approximation is made to arrive at the dispersion relation for the form factor $F(Q)$. Both approximations split at the very end. To begin, we substitute the IR for propagators and vertices, and by  changing the ordering 
  of integrations we perform momentum integrations following standard procedures known from perturbation theory.

 Thus, after  the   Feynman paramatrization, we can perform integration over the momentum exactly in a way known  for evaluation of Feynman integrals in  perturbation theory. 
 As we are not in perturbation theory, we remain with number of integrals over the weight functions of all integral representations, 
 as well as all those with  three  new auxiliary integrals. For the later we use the Feynman variable $x$ to mach the denominators of two IRs for the BSE vertices,
 then we use the variable $y$  to match the result  with the denominator of the quark propagator, which connects these two vertices, and at the end we will use the variable $t$ 
 in order to match rhe result of previous matching with the denominator of the IR for the quark-photon vertex.

The above described steps  [for the entire  matrix element ${\cal J}^{\mu}(Q|GT)$] read explicitly
 \bea
 \label{ulanbatar}
{\cal J}^{\mu}(Q)&=&-i 2N_c\int \frac{d^4 k}{(2\pi)^4} \int_s   \frac{\Gamma(5) y t^2(1-t)  U^{\mu}}{[(k+l)^2+J]^5} \, ,
 \\
U^{\mu}&=&4\rho_v(\gamma)\rho_v(\omega)\left[\left(\omega-k^2+\frac{Q^2}{4}\right) (p^{\mu}-k^{\nu})+2k^{\mu}
(k.p-k^2)+\frac{1}{2}Q^{\mu}k.Q\right]
\nn \\
&&8\rho_s(\gamma)\rho_s(\omega) k^{\mu}
\nn \\
J&=&-l^2+\frac{Q^2}{4}(1-t)-\om(1-t)+p^2(1-y)t-\gamma(1-y)t+\left[\frac{p^2}{4}+\frac{Q^2}{16}-ax-b(1-x)\right]yt \, ,
\nn \\
l^2&=&\frac{Q^2}{4} o^2+p^2(-1+\frac{y}{2})^2t^2  \, ,
\nn \\
o&=&z(1-t)-\frac{1-2x}{2}yt \,
\label{ocko}
\eea
where we have used some shorthand notations: mainly we have also factorized the weight functions $\rho_{\pi}$ of the pion vertex functions into the overall measure, for this we use the following abbreviation 
\be
\int_s =\int_0^1 dx dy dt \int_{0}^{\infty} d{\om} \int_{-1}^{1} dz \int da \int db \rho_{\pi}(a)\rho_{\pi}(b).
\ee
but we  omit all trivial terms which were proportional to the product of external momenta $Q.p=0$. However we keep  $p^2$ variable
  for the purpose of easier tracking of the presented derivation.
 We will use the fact that  pions are on-shell, i.e. the equation  $p^2=m_{\pi}^2-Q^2/4$ from the following lines.

For the  purpose of  integration over the momentum  we perform the standard shift 
$k\rightarrow k-l$ where $l=\frac{Q}{2} o+p(-1+y/2)t$, with the polynomial function $o$ defined by (\ref{ocko})  
\be
o=z(1-t)+\frac{1-x}{2}yt \, ,
\ee
which after  the integration over the momentum  provides the  nontrivial part of our matrix element
in the form
\be
{\cal{J}}^{\mu}(p,Q)=F(Q^2)p^{\mu} \,  ,
\ee
where the pion form factor $F$ is proportional to the following expression
\bea
F(Q^2)&=&\frac{2N_c}{(4\pi)^2)} \int_s y t^2(1-t) \frac{4\rho_v(\gamma)\rho_v(\omega)}{(4\pi)^2 J^3}\left[(1-f)\omega +f(-\frac{Q^2}{4}(1+o^2)
+2p^2 f-p^2 f^2)\right]
\nn \\
&+&\frac{4\rho_s(\gamma)\rho_s(\omega)}{(4\pi)^2 J^2}
+\frac{4\rho_v(\gamma)\rho_v(\omega)}{(4\pi)^2 J^3}
(1-3f) \, ,
\label{after}
\eea
where we have labeled 
\be
f=(1-\frac{y}{2})t \,  .
\ee

The individual   prefactors in  Eq. (\ref{after}) follow from the standard evaluation performed   
in  Euclidean space, although  we come back 
to the Minkowski metric convention immediately. We just remind the reader
with the  example 
\be
-i\int \frac{d^4k}{(2\pi)^4} \frac{\Gamma(5) k^2}{(k^2+J)^5}=\frac{2}{(4\pi)^2 J^2}\, .
\ee

Note here that for purpose of consistency,  the variable $p^2$ was also kept Euclidean for a while, 
 and the on- shell condition $p^2=m_{\pi}^2-Q^2/4$ is imposed only afterwords.
 
 For positive timelike $Q^2$ the real part of  denominator $J$ pass zero value and albeit not written explicitly, the presence of infinitesimal 
 Feynman imaginary part is assumed.

In what follows it is convenient to split  the denominator $J$, getting
 \bea
J&=&\frac{Q^2}{4}\square-\Delta
\nn \\
\Delta&=&m_{\pi}^2[(1-\frac{3}{4}y)t-f^2]
-\omega(1-t)-\gamma(1-y)t-(ax+b(1-x))yt
\nn \\
\square&=&-o^2+1-t-(1-y)t+f^2 \, .
\eea

In  Eq. (\ref{after}) we do not write  trivial terms, including also those, which are proportional linearly
 to the variable $o$. These terms are zero as can be inspected by   the substitutions $z\rightarrow -z$ and $x\rightarrow 1-x$ 
 with simultaneous  interchange of the pion spectral function arguments $a\rightarrow b$. In this way one 
 gets the identical  expression for the appropriate contributions to the  form factor, 
 but with opposite sign; hence, it is  zero. This, together with on-shell condition $p.Q=0$, causes that term proportional to total momentum
 $Q^{\mu}$ to be absent for each diagram individually and the matrix element has an identical Lorentz structure  
 to the charged point like scalar particle.

 Before evaluating singular and hence more complicated Minkowski expressions we 
derive the formula suited for the numerical integrations for spacelike momentum $Q$.
For this purpose we integrate  over the variable  $z$ variable analytically.
To proceed furthermore, we use "gauge technique" trick  again and make linearization in $\rho_{\pi}$, which allows us
to  reduce the number of integrations further. As a consequence $x a +(1-x)b\rightarrow b$
and the integration over the variable $x$ can be done  analytically in closed form.
For purpose of numeric,  the same is done for the product of the 
quark spectral function integrals, where after matching 
by the virtue of  gauge technique linearization we make linearization in $\rho_v$ ($\rho_s$) such that 
$\om(1-t)-\gamma(1-y)t\rightarrow \gamma(1-yt)$. In what follows, we will write $\tilde{\gamma}=
\gamma(1-yt)+b yt$

There are only two  necessary integrals  for  evaluation of  the function 
$F$ for the spacelike momentum 
$Q^2_E=-Q^2$. The first we show here
\bea
\int_{-1}^{1}dz \int_0^1 dx J^{-3}&=&
\frac{\theta(a)}{\frac{Q_E^2}{4} (1-t)t y } D_x\left[\frac{1}{4a(x^2+a)}-
\frac{3x \arctan[\frac{x}{\sqrt{a}}]}{4a^{5/2}}\right]  
\nn \\
&+&\frac{\theta(-a)}{\frac{Q_E^2}{4} (1-t)t y } D_x\left[\frac{1}{4a(x^2+a)}-
\frac{3x \tanh^{-1}[\frac{x}{\sqrt{-a}}]}{4(-a)^{5/2}}\right]
\eea
with the function $a_d$ defined as
\be
a_d=-\frac{Q_E^2}{4}\left[(1-t)-(1-y)t+f^2\right]+
(1-3/4y-f^2)m_{\pi}^2-\tilde{\gamma}\, .
\ee

The second required integral reads
\be
\int_{-1}^{1}dz \int_0^1 dx J^{-2}=
\frac{-\theta(a)}{\frac{Q_E^2}{4}(1-t)t y } D_x
\left[x\frac{ \arctan[\frac{x}{\sqrt{a}}]}{a^{3/2}}\right]  
-\frac{\theta(-a)}{\frac{Q_E^2}{4}(1-t)t y } D_x 
\left[x \frac{ \tanh^{-1}[\frac{x}{\sqrt{-a}}]}{(-a)^{3/2}}\right]
\ee
where we have introduced abbreviations
\bea
D_x [h(x)]&=&h(x_u)-h(x_d) \, ;
\nn \\
x_u=Q_E/2[1-t-yt/2]\, \, &;& x_d=Q_E/2[1-t+yt/2]\, ;
\eea
for some function $[h(x)]$.

Using the above integrals in  Eq. (\ref{after}) constitutes the final  expression that we have used for the numerical
evaluation for the spacelike value of $Q^2$.

\subsection{Derivation of Dispersion Relation}

 The final  expression for $F$ based on formulas  derived above is  still not yet in a form suited for numerical evaluation in the region of  Minkowski momentum $Q^2>0$.
  Recalling the presence of the small Feynman factor $\ep$ , the log in inverse hyperbolical tangents as well as the function $a$ is badly singular  near the real axis of momentum $Q$ and the expression  is 
 numerically ill. Since complete analytical integration is still out of our reach, we make further simplification. 
For this purpose we go back into the expression and ignore the presence of the  $o^2$ term in the integrand
which allows the conversion of all terms into the desired dispersion relation.
We show the derivation   for most singular   $1/J^3$ term in the Eq. (\ref{after}) the conversion of other terms is straightforward within the method used. 

Ignoring $o^2$  terms as well as ignoring small 
terms proportional to $m_{\pi}$ we  can integrate over the 
variables $x$ and $z$. The result simply means to replace the integration symbols $\int dx dz $ with a factor $2$. 
The remaining relevant integral   we need to evaluate reads
\be
\int_0^1 dt \int_0^1 dy \frac{(1-y/2)^2 y t^2 (1-t)}{J^3} \, ,
\label{sofar}
\ee
where the denominator reduces as
\be
J=\square \frac{Q^2}{4}-\tilde{\gamma} \, ,
\ee

To proceed further, we perform the last  linearization 
\be
\int d \gamma \rho_{v,s}(\gamma) \int d b\rho_{\pi}(b)\rightarrow  \int d \gamma \tilde{\rho}_{v,s}(\gamma) \, ,
\label{linear}
\ee
where  we assume new functions $\tilde{\rho}$ on the rhs.  (\ref{linear}) are  such that resulting form factor  $F$ remains unchanged when taking 
$\tilde{\gamma}\rightarrow \gamma $ in the denominator, i.e. from now
 \bea
J&=&\square \frac{Q^2}{4}-\gamma \, ,
\nn \\
\square&=& 1-t-(1-y)t+(1-y/2)^2 t^2 \, .
\eea 
and we also assume $\tilde{\rho}\simeq \rho+\delta_{\rho}$ with the function $\delta_{\rho}$  representing  corrections.

In addition we introduce the unit in the form
\be
1=\int_0^{\infty} d \alpha \delta(\alpha-\gamma/\square)
\ee
into the expression (\ref{sofar}) and integrate over the variable $t$. After that   we get for (\ref{sofar}) the following expression
\be
\int_0^1 dy \int_0^{\infty} d \alpha  \frac{\alpha}{\gamma^2}
 \frac{ yt_-^2(1-t_-)\theta(t_-)\theta(1-t_-)}{2 (1-(1-y/2)t_-)[\frac{Q^2}{4}-\alpha+\ep]^3}
 \ee
 where $t_-$ is the root of the equation $\square \alpha-\gamma=0$. Explicitly it reads
 \be
 t_-=\frac{1-\sqrt{\gamma/\alpha}}{1-y/2}
 \ee
 noting that  since  $\alpha>0$, the step function can be equivalently taken as $\theta(\alpha-\gamma)\theta(4\frac{\gamma}{y^2}-\alpha)$.
 Note that the contribution from  the second root $t_+=1+/..$ is trivial , since $t_+>1$, being thus always outside of the  interval for the original integral variable $y$.

Let us change the ordering of the integrations and integrate over the variable $y$.  Theta functions presented in the kernel  imply
 \be
 \int_0^1 dy \int_0^{\infty} d\alpha \rightarrow 
 \int_{\gamma}^{\infty} d\alpha \int_0^{2\sqrt{\frac{\gamma}{\alpha}}} dy
 \ee
 After the integration we get 
 \be
 \int_0^{\infty} d\alpha \frac{4}{\gamma} \frac{\theta(1-\sqrt{\frac{\gamma}{\alpha}})\theta(\alpha-\gamma)}
 {[\frac{Q^2}{4}-\alpha+\ep]^3}
\left [-2-(2\sqrt{\frac{\alpha}{\gamma}}-1) ln(1-\sqrt{\frac{\gamma}{\alpha}})\right](\sqrt{\frac{\alpha}{\gamma}}-1)^2
 \ee
 
 After that,  we perform double per partes integration with respect to the variable $\alpha$, such that we get the desired dispersion relation
  \bea
F(Q^2)&=& \int_{0}^{\infty} d\alpha \frac{g(\alpha)}{[\frac{Q^2}{4}-\alpha+\ep]} 
\nn \\
g(\alpha)&=&\frac{4 N_c}{(4\pi^2)}\int_0^{\alpha} d\gamma \frac{2\hat{\rho_v}(\gamma)}{\gamma}K(\alpha,\gamma)+... 
\nn \\
K(\alpha,\gamma)&=&\frac{B A^2+A+B-1/2}{2\alpha^{3/2}\gamma^{1/2}}
 -\frac{2+A.B}{\alpha\gamma}-\frac{B}{\alpha^{1/2}\gamma^{3/2}}
 +\frac{1}{2\alpha^2}     
 \label{ADR}
\eea
 where 
 \be 
A=1-\sqrt{\frac{\alpha}{\gamma}} \, \, ;  \, \, B=ln(1-\sqrt{\frac{\gamma}{\alpha}})
\ee
and where dots represent  remaining and not shown contributions  (stemming also  from the integration over the  function $J^{-2}$, we found  these terms  can be safely 
neglected in the approximation employed here).

 Our approximation leads to some  systematical error: it smoothly overestimates the form factor at medium timelike $Q^2$ and 
 the dispersion relation does not provide correct  form factor for $|Q^2|>2 GeV^2$, hence we call the form factor calculated 
 on the relation (\ref{ADR}) the  Approximated Dispersion Relation  result.
 Nevertheless,  it offers reasonable comparison with approximation derived in the previous section.
 Hence, we guess that our ADR does not cripple
  the  function $F$ bellow  1GeV too much,  keeping  the shape of $\rho$ meson resonance not distorted much.


%

\begin{thebibliography}{00}

\bibitem{I} 
A. Aloisio, et al (KLOE Collaboration), Phys. Lett. B 606, 12 (2005);  V. M. Aulchenko, et al (CMD-2 Collaboration), JETP Lett. 82, 743 (2005); PismaZh. Eksp. Teor. Fiz. 82, 841
(2005);  M.N.Achasov,et al, J. Exp. Theor. Phys. 103, 380 (2006); Zh. Eksp. Teor. Fiz. 130, 437 (2006); J. P. Lees (BABAR Collaboration), Phys. Rev. D {\bf 86}, 032013 (2012);  M. Ablikim, (BESIII Collaboration) Phys. Lett. B753, 629 (2016).

\bibitem{COPETH1997}
H. B. O’Connell, B. C. Pearce, A. W. Thomas and
A. G. Williams, Prog. Part. Nucl. Phys. 39 (1997),
201-252 doi:10.1016/S0146-6410(97)00044-6.

\bibitem{data2006}
R. R. Akhmetsin {\it et al.}, JETP Lett B {\bf 84}, 413 (2006). 

\bibitem{tade2007}
V. Tadevosyan {\it et al.}, Phys. Rev. D{\bf 75}, 055205 (2007).

\bibitem{aubert2007}
B. Aubert {\it et al.}, Phys. Rev. Lett. {\bf 103}, 231801 (2009).

\bibitem{data2007}
R. R. Akhmetsin {\it et al.}, Phys. Lett. B {\bf 648}, 28 (2007)

\bibitem{BABAR2012}
J. P. Lees {\it et al.}, (BABAR Collaboration), Phys. Rev. D{\bf 86}, 032013 (2012).

\bibitem{CHZH1977}
V. Chernyak and A.R. Zhitnisky, JETP Lett. {\bf 43}, 510 (1977); Sov. J Nucl. Phys. {\bf 31}, 544 (1980).  

\bibitem{FAJA1979}
G. Farrar and D. Jackson, Phys. Rev. Lett. {\bf 43}, 246 (1979).
%
\bibitem{LEBR1979}
G. P. Lepage and S. J. Brodsky, Phys. Lett. {\bf 87 B}, 359 (1979);  Phys. Rev. D {\bf 22}, 2157 (1980).
%
\bibitem{EFRA1980}
A. V. Efremov and A. V. Radyushkin, Phys. Lett. {\bf94} B, 245 (1980).  
%
\bibitem{LE2002}
H. Leutwyler, Electromagnetic form factor of the pion,
Continuous Advances in QCD 2002/Arkadyfest, 23 (2002)


\bibitem{GALE1985}
J. Gasser and H. Leutwyler, Nucl. Phys. {\bf B 250}, 517 (1985).

\bibitem{CFU1996}
G. Colangelo, M. Finkemeier and R. Urech, Phys. Rev. D {\bf 54}, 4430 (1996).

\bibitem{BCT1998}
J. Bijnens, G. Colangelo and P. Talavera, JHEP 9805, 014 (1998).

\bibitem{BITA2002}
J. Bijnens and P. Talavera, JHEP 0203,046 (2002).

%
\bibitem{BIJDHO2003}
 J. Bijnens and P. Dhonte, JHEP {\bf 0310}, 061 (2003).

\bibitem{KAROL2012}
K. Kampf, Nuc. Phys. {\bf B 234}, 299 (2013) .

\bibitem{DGL1990}
 J. F. Donoghue, J. Gasser and H. Leutwyler, Nucl. Phys. {\bf B 343}, 341 (1990).

\bibitem{GAME1991}
J. Gasser and U. Meißner, Nucl. Phys. {\bf B 357}, 90 (1991).

\bibitem{DN1997}
J. F. Donoghue and E. S. Na, Phys. Rev. D {\bf 56}, 7073 (1997).

\bibitem{CA471}
I. Caprini, Eur. Phys. J. C {\bf 13}, 471 (2000).

\bibitem{PIPO2001}
A. Pich and J. Portoles, Phys. Rev. D {\bf 63}, 093005 (2001).

\bibitem{OOP2001}
J. A. Oller, E. Oset and J. E. Palomar, Phys. Rev. D {\bf 63}, 114009 (2001).

\bibitem{TY2002}
 J. F. De Troconiz and F. J. Yndurain, Phys. Rev. {\bf D 65}, 093001 (2002).


%
\bibitem{DGKST2015}
D. Djukanovic, J. Gegelia, A. Keller, S. Scherer, L. Tiator, Phys. Lett.B {\bf 742}, 55-60 (2015). 
\bibitem{MT99}
P. Maris, P.C. Tandy, Phys. Rev. C {\bf 61}, 045202 (2000).
%
\bibitem{MATA200a}
P. Maris and P. C. Tandy, Phys. Rev. {\bf  61}, 045202 (2000).
%
\bibitem{MATA200b}
P. Maris and P. C. Tandy,  Phys. Rev. C {\bf  62}, 055204 (2000).
%
\bibitem{kacka2005}
P. Maris,  P.C. Tandy, Phys. Rev. C {\bf 62}, 055204 (2000). 
%
\bibitem{CHA2013}
 Chang, L. et al. Phys. Rev. Lett. {\bf 111}, 141802  (2013) .
%
\bibitem{RCBRG2016}
 K. Raya, et al., Phys. Rev. D {\bf 93}, 074017 (2016). 
 %
\bibitem{CDCL2017}
 J. Chen, M. Ding, L. Chang, Y. Liu, Phys. Rev. D {\bf  95}, 016010 (2017).
%
\bibitem{EFWW2017}
G. Eichmann, Ch. S. Fischer, E. Weil, R. Williams, Phys. Lett. B {\bf 774}, 425-429 (2017).
%
\bibitem{WEFW2017}
E. Weil, G. Eichmann, C. S. Fischer, R. Williams , Phys. Rev. D {\bf 96}, 014021  (2017).
%
\bibitem{DRBBCCR2019}
M. Ding, K. Raya, A. Bashir, D. Binosi, L. Chang, M. Chen, C. D. Roberts, Phys. Rev. D {\bf 99}, 014014 (2019).   

\bibitem{YPNFS2021}
E. Ydrefors, W.de Paula , J.H.A. Nogueira, T.Frederico, G.Salme, Phys. Lett. B {\bf 820}, 136494 (2021).

\bibitem{VHP2020}
V. Sauli,    Few Body Syst. {\bf 61}, 3, 23 (2020).

\bibitem{TPF2020}
 V. Sauli,   Phys. Rev. D {\bf  102}, 014049 (2020). 
%
\bibitem{LSA2021}
 A.S.M. Lopez, H. Sanchis-Alepuz, R. Alkofer,  Phys. Rev. D {\bf  103}, 116006 (2021). 
\bibitem{KKW1996}
 F. Klingl, N. Kaiser, W. Weise, Z. Phys. A{\bf 356}, 193 (1996).
%
\bibitem{LL2009}
 S. Leupold, M. F. M. Lutz,   Eur. Phys. J. A {\bf 39}, 205 (2009).
 %
\bibitem{B2009}
T. Bauer, AIP Conf.Proc. 1432, 1, 269-272 (2012);  arXiv:1107.5506. 
%
\bibitem{TSLE2013}
 C. Terschlüsen, B. Strandber, S. Leupold, F. Eichstädt,   Eur. Phys. J. A {\bf 49}, 116 (2013).
%
\bibitem{BEBUDULI2011}
M. Belicka, S. Dubnicka, A. Z. Dubnickova, and A. Liptaj, Phys. Rev. C{\bf 83}, 028201 (2011).
%
 \bibitem{HKLNS2014}
M. Hoferichter, B. Kubis, S. Leupold, F. Niecknig, and S. P. Schneider, Eur. Phys. Jour. C{\bf 74}, 3180 (2014).
%
%
\bibitem{GOGUAD2012}
M. Gorchtein, P. Guo,  and A. P. Szczepaniak, Phys. Rev. C {\bf 86}, 015205 (2012).
%
\bibitem{SORO2019}
 S. Gonzalez-Sol\'{i}s, P. Roig; Eur. Phys. J. C {\bf 79}, 436 (2019).

%
\bibitem{MANDEL}
S. Mandelstam, Proc. R. Soc. Lond. A{\bf 233}, 248  (1955).
%
\bibitem{BHKRT2005}
M. S. Bhagwat, A. H\''{o}ll, A. Krassnigg, C. D. Roberts, and P. C. Tandy, Phys. Rev. C {\bf  70}, 035205 (2004).
%
%
\bibitem{MR1997}
 P. Maris, C.D. Roberts, Phys. Rev. C {\bf 56}, 3369 (1997).
%
\bibitem{MT1999}
 P. Maris, P.C. Tandy,  Phys. Rev. C {\bf 60}, 055214 (1999).
 %
\bibitem{MATA2002}
P. Maris and P. C. Tandy, Phys Rev. C {\bf 65}, 045211 (2002).
%
\bibitem{JIMA2001}
C.-R. Ji and P. Maris, Phys Rev. D {\bf 64}, 014032 (2003).
%
\bibitem{BPT2003}
M. S. Bhagwat, M. A. Pichowski, and P. C. Tandy, Phys. Rev. D {\bf  67}, 054019 (2003).
%
\bibitem{JAMATA2003}
D. Jarecke, P. Maris, and P.C. Tandy, Phys. Rev. C {\bf 67}, 035202 (2003).
%
\bibitem{HKRMW2005}
A. Hoell, A. Krassnigg, P. Maris, C. D. Roberts, and S. V. Wright, Phys. Rev. C{\bf 71}, 065204 (2005). 
%
 \bibitem{EACK2008}
 G. Eichmann, R. Alkofer, I.C. Cloet, A. Krassnigg, C.D. Roberts, Phys. Rev. C {\bf 77}, 042202(R) (2008).
%
\bibitem{BEIVRO2011}
B. El-Bennich, M. A. Ivanov, and C. D. Roberts Phys. Rev. C {\bf  83}, 025205 (2011).
%
\bibitem{dresden2}
S. M. Dorkin, L. P. Kaptari, T. Hilger, and B. Kampfer, Phys. Rev. C {\bf  89}, 034005 (2014).
%
\bibitem{HGK2015}
T. Hilger, M. Gomez-Rocha, A. Krassnigg,  Phys. Rev. D{\bf 91},  114004  (2015).
%
\bibitem{HPRG2015}
T. Hilger, C. Popovici, M. Gomez-Rocha, A. Krassnigg ,  Phys. Rev. D{\bf 91}, 034013 (2015).
%
\bibitem{dresden1}
 S. M. Dorkin, L. P. Kaptari, and B. Kampfer, Phys. Rev. C {\bf 91} , 055201 (2015).
%
\bibitem{HGKL2017}
T. Hilger, M. Gómez-Rocha, A. Krassnigg, W. Lucha,  Eur. Phys. J. A {\bf 53}, 213 (2017). 
%
\bibitem{QRS2019}
S.-X. Qin, C. D. Roberts and S. M. Schmidt, Few-Body Syst, {\bf 60},26 (2019).
%
\bibitem{Sauli2012}
V. Sauli, J. Phys. G {\bf 39} 035003 (2012).
%
\bibitem{STCA1990}
  S. J.  Stainsby and R.T.  Cahill, Phys. Lett. {\bf A 146} 9, 467 (1990).
%
%
\bibitem{WI2017}
A. Windisch. Phys. Rev. C {\bf 95}, 045204 (2017).
%
\bibitem{VS2014}
V. Sauli,  Phys. Rev. D {\bf 90},  016005  (2014).
%
\bibitem{SAJPSI}
V. Sauli, Phys. Rev. D {\bf 86}, 096004 (2012). 
\bibitem{Naka1961}
N. Nakanishi,  Prog. Theor. Phys. {\bf 26} , 375 (1961). 
%
\bibitem{SA2008}
 V. Sauli,  J. Phys. G {\bf 35}, 035005 (2008). 
%
\bibitem{MMFP2022}
R. M. Moita, J. P.B. C.de Melo, T. Frederico , W. de Paula ,
Rev. Mex. Fis. Suppl. 3, 0308089 (2022); arXiv:2202.10959.
%
\bibitem{ZRC2021}
Z. Zhu, K. Raya, L. Chang, Phys. Rev. D {\bf 103}, 034005 (2021).

\bibitem{GT1963}
 A. Salam, Phys. Rev. {\bf 130}, 1287 (1963).
%
\bibitem{GT1964}
 J. Strathdee, Phys. Rev.{\bf 135}, 1428 (1964).
%
\bibitem{GT1977}
 R. Delbourgo, P. C. West, J. Phys. A {\bf 10}, 1049 (1977).
%
\bibitem{GTII}
 R. Delbourgo, P.C. West, Phys. Lett. B {\bf 72}, 96 (1977).
%
\bibitem{AB2003}
E. R. Arriola and W. Broniowski, Phys. Rev., D{\bf 67}, 074021 (2003).
%
\bibitem{MRSB2004}
 E. Megias, E. R. Arriola, L. L. Salcedo, and W. Broniowski, Phys. Rev.
D{\bf 70}, 034031 (2004).
%
\bibitem{ABG2007}
E. R. Arriola, W. Broniowski, and B. Golli, Phys. Rev. D{\bf76}, 014008 (2007).
%
\bibitem{oser2017}
V. Sauli , , EPJ Web Conf. 179, 01021 (2018); ArXiv 1708.03616. 
%
\bibitem{SAAD2003}
    V. Sauli, J. Adam, Jr. ; Phys. Rev. {\bf D 67}, 085007 (2003).
%
\bibitem{KUSIWI1997}
K. Kusaka, K. Simpson , A.G. Williams, Phys. Rev. {\bf D 56}, 5071 (1997).
%
\bibitem{SAJPA2003}
V. Sauli,  J. Phys. A{\bf 36}, 8703-8722 (2003).
%
\bibitem{SAJHP2003}
V. Sauli,  JHEP  {\bf 02}, 001 (2003).
\bibitem{bebek1978}
C. J. Bebek, at all..  Phys. Rev. {\bf D 17}, 1693 (1978).

\bibitem{CEA}
C. N. Brown, at. all, Phys. Rev. D 8, 92 (1973).

\bibitem{cornell1}
C. J. Bebek, at all., Phys. Bev. {\bf D 9}, 1229 (1974).

\bibitem{cornell2}
C. J. Bebek, at all. , Phys. Bev. {\bf D 13}, 25 (1976).




\end{thebibliography}
\end{document}